\documentclass[aps,prd,onecolumn,groupedaddress,12pt,showpacs,showkeys]{revtex4-1}

\usepackage{amsmath,amssymb,MnSymbol}
\usepackage{graphicx,slashed}

\newcommand{\bs}{\boldsymbol}{}

\newcommand{\eq}{\begin{eqnarray}}
\newcommand{\en}{\end{eqnarray}}

\newcommand{\p}{\partial}
\newcommand{\m}{\mathbf}
\newcommand{\mb}{\boldsymbol}
\newcommand{\lp}{\stackrel{\leftarrow}{\partial}}
\newcommand{\rp}{\stackrel{\rightarrow}{\partial}}
\newcommand{\lnabla}{\stackrel{\leftarrow}{\nabla}}
\newcommand{\rnabla}{\stackrel{\rightarrow}{\nabla}}

\begin{document}


\title{Nucleon in a periodic magnetic field: finite-volume aspects}


\author{Andria Agadjanov$^{a}$, Ulf-G. Mei{\ss}ner$^{a,b,c}$ and Akaki Rusetsky$^a$}
\affiliation{$^a$Helmholtz-Institut f\"ur Strahlen- und Kernphysik (Theorie) and
Bethe Center for Theoretical Physics, Universit\"at Bonn, D-53115 Bonn, Germany}
\affiliation{$^b$Institute for Advanced Simulation (IAS-4), Institut f\"ur Kernphysik 
	(IKP-3) and  J\"ulich Center for Hadron Physics,
	Forschungszentrum J\"ulich, D-52425 J\"ulich, Germany}
\affiliation{$^c$Ivane Javakhishvili Tbilisi State University, 0186 Tbilisi, Georgia}

\date{\today}

\begin{abstract}
  The paper presents an extension and a refinement of our previous work
  on the extraction of the doubly virtual forward Compton scattering amplitude
  on the lattice by using the background field technique~\cite{Agadjanov:2016cjc}.
  The zero frequency limit for the periodic background field is discussed,
  in which the well-known result is reproduced. Further, an upper limit for the magnitude
  of the external field is established for which the perturbative treatment is still possible.
  Finally, the framework is set for the
  evaluation of the finite-volume corrections allowing for the analysis  of upcoming lattice results.

\end{abstract}


\maketitle

\section{Introduction}

Using the background field technique on the lattice for the extraction of various
hadronic observables has proven to be extremely efficient. As examples we
mention the measurement of magnetic moments, polarizabilities and axial-vector
matrix elements of baryons and light nuclei
in constant background fields~\cite{Savage:2016kon,Chang:2015qxa,Beane:2014ora,Beane:2015yha,Chambers:2015bka,Chambers:2014qaa,Chang:2015qxa}.
Moreover, non-uniform background fields
have been used for the calculation of the hadronic vacuum polarization tensor,
hadronic form-factors and the nucleon structure functions, as the fields, which are periodic in space,
allow one to measure  current matrix elements
at a given non-zero three-momentum transfer~\cite{Bali:2015msa,Chambers:2017tuf,Chambers:2017dov}.
Different scenarios for implementing periodic background fields in lattice QCD
calculations are considered in Ref.~\cite{Davoudi:2015cba}.

In Ref.~\cite{Agadjanov:2016cjc} we described a framework, based on the background field method, 
which enables one to extract the doubly-virtual forward Compton scattering amplitude from
lattice QCD calculations (note that later, in Ref.~\cite{Chambers:2017dov}, a very similar formula was given without a derivation, see Eq.~(12) in that paper).
Low-energy Compton scattering plays an
indispensable role in probing the electromagnetic structure of hadrons (for recent work,
see Ref.~\cite{Hagelstein:2015egb}). For example,
it enters the expression for the proton-neutron
electromagnetic mass difference~\cite{Cottingham:1963zz}, as well as the expression
for the Lamb shift of the muonic hydrogen, which is used to extract the value of the
proton radius (see, e.g.~Ref.~\cite{Carlson:2011zd}).
In this paper, we in particular focus on the relevant spin-independent 
invariant amplitudes, denoted as $T_1$ and $T_2$, respectively. The experimental data on the 
structure functions completely determine the amplitude $T_2$ through  dispersion
relations. The fixed-$q^2$ dispersion
relation for the amplitude $T_1$, however, requires a subtraction. Thus,
the subtraction function $S_1(q^2)\equiv T_1(0,q^2)$ remains the only input
in the calculations, which is not fixed by experimental data. 
Its elastic part is essentially given  by the Born terms, but the inelastic piece is
known only at the real photon point $q^2=0$ (the low-energy theorem, see, e.g.,
Ref.~\cite{Gasser:2015dwa}).

In the past, there have been attempts to model the subtraction function, using
phenomenological
parameterizations~\cite{WalkerLoud:2012bg,Erben:2014hza,Cushman:2018zza}.
However, this type of approach inherently contains a systematic error, which is very hard
to control. Further, the subtraction function can be extracted from the Compton scattering
amplitude, calculated in the low-energy EFT of
QCD~\cite{Bernard:2002pw,Peset:2014jxa,Bernard:2012hb}. However, here the difficult question about
the convergence of the chiral expansion arises, namely, up to which value of $q^2$
the results of the chiral expansion can be trusted.
Recently, the authors of Ref.~\cite{Gasser:2015dwa} have been able to determine
the subtraction function by invoking  the so-called Reggeon dominance hypothesis,
considered first in Ref.~\cite{Gasser:1974wd}. In particular, it is assumed
that the forward Compton scattering amplitude does not contain any fixed pole.
In Regge theory, such a pole  generates an energy-independent contribution to
the amplitude (such as, e.g., local two-photon couplings in scalar QED).
If the fixed poles are present, the subtraction function, in general, deviates
from the predicted one. In some cases, e.g., the $q^2$-independent fixed
pole~\cite{Brodsky:2008qu}, the behavior of the $S_1(q^2)$ can be also predicted
(see Ref.~\cite{Gorchtein:2013yga}), and is different from the one
calculated in the absence of the fixed pole.

Hence,
lattice QCD provides a model-independent approach to the verification of the
Reggeon dominance hypothesis. The question whether there is a fixed pole in
the Compton scattering is of conceptual interest and is still open.

In Ref.~\cite{Agadjanov:2016cjc}, we considered the case of a
nucleon placed in a static periodic magnetic field
${\bf B}=(0,0,B^3)$ with $B^3=-B\cos(\boldsymbol{\omega}{\bf x})$ and 
$\boldsymbol{\omega}=(0,\omega,0)$. It has been
shown that, for $\omega\neq 0$, the measurement of the spin-averaged energy
shift of the nucleon in this field allows one to extract the value of the subtraction
function $S_1(-\boldsymbol{\omega}^2)$ at non-zero values of $q^2=-\boldsymbol{\omega}^2$. 
The relation between these two quantities takes the form:
\eq
\delta E=\frac{(eB)^2}{4m}\, S_1(-\boldsymbol{\omega}^2)\, .
\en
Here, $\delta E$ denotes the spin-averaged energy shift, and $m$ stands for the nucleon mass.

The result, obtained in Ref.~\cite{Agadjanov:2016cjc}, still leaves
room for improvement. In particular, one should find the answer to the following
questions:
\begin{itemize}

\item[i)] In the limit $\omega\to 0$, we arrive at the case of a
  constant magnetic field. This case is studied very well, both analytically and numerically.
  However, our expressions become singular in this limit, contradicting the expectations.
  One needs to understand how this limit can be approached smoothly.

\item[ii)]
  Our approach relies
  on a perturbative expansion of the energy shift $\delta E$ in the external field
  strength $B$. What is the radius of the convergence of this expansion? Note that, for
  example, in the zero-frequency limit  $\omega\to 0$, the radius is
  equal to zero in case of a charged particle, since the Landau levels are formed
  for any value of $B$. In Ref.~\cite{Agadjanov:2016cjc}, using heuristic arguments,
for a given value of $\omega$,  we gave a very rough estimate of the maximal value of $B$, for which the perturbative
  expansion should still work. These arguments should
  be refined  in order to obtain a more reliable result.
  
\item[iii)] Our expressions were obtained in the infinite-volume limit. The issue of the
  finite-volume corrections in the presence of the external fields is a rather subtle one, 
  since gauge-invariant non-local objects (Wilson lines) can be formed in a finite volume.
  For this reason, it is mandatory to re-formulate the problem in a finite volume
  from the beginning
  and to give a consistent interpretation of the finite-volume result it terms of the
  subtraction
  function, defined in the infinite volume.

  \end{itemize}

  The aim of the present paper is to answer the questions given above. The plan of the
  paper is as follows. In section~\ref{sec:notions} we give a collection
  of basic definitions and discuss two different implementations of the external field on the
  lattice. In sections~\ref{sec:NR} and \ref{sec:rel}, we give two alternative
  derivations of the energy shift formula in the periodic external field, based on the
  matching to the non-relativistic EFT, as well as the direct derivation within the relativistic
  framework. Both settings are complementary to each other. For example, the first
  derivation is more intuitive and uses ordinary
  quantum-mechanical Rayleigh-Schr\"odinger perturbation theory for the energy levels.
  In particular,
  the zero-frequency limit $\omega\to 0$ as well as the issues
  related to the convergence of the perturbative expansion can be considered
  more easily in this formulation.
  By contrast, the relativistic formulation allows one to investigate the exponentially
  suppressed finite-volume corrections in a direct manner. 
  For completeness, in the appendix we give yet another derivation of the
  expression for the energy shift, considering the behavior of the nucleon two-point
  function at large time separations.

\section{Definitions and setup}
\label{sec:notions}
  
\subsection{Basic definitions}

The matrix element of the electromagnetic current between one-nucleon states is given by
\begin{equation}\label{current matrix}
\langle p',s'|j^\mu(0) |p,s\rangle = \bar{u}(p',s') \Big\{ \gamma^\mu F_1(q^2)+ i \sigma^{\mu\nu}F_2(q^2)\frac{q_\nu}{2m}\Big\}u(p,s)\, .
\end{equation}
Here, $j^\mu(x)$  is the electromagnetic current, and 
$q=p'-p$, and $p\,(p')$ and $s\,(s')$ are the four-momenta and spin projections of the
initial (final) nucleon, respectively.
Further, $F_1$ and $F_2$ denote the Dirac and Pauli
form factors. The Sachs form factors are defined by
\begin{equation}
  G_E(q^2)=F_1(q^2)+\frac{q^2}{4m^2}F_2(q^2)\, ,\quad\quad
  G_M(q^2)=F_1(q^2)+F_2(q^2)\, .
\end{equation}
The Dirac spinors are normalized as $\bar{u}(p,s')u(p,s)=2m\delta_{s's}$.

The Compton tensor is defined as:
\begin{equation}\label{Compton tensor}
T^{\mu\nu}(p',s';p,s;q)=\frac{i}{2}\int d^4x e^{i q\cdot x}\langle p',s'|Tj^\mu(x)j^\nu(0)|p,s\rangle,
\end{equation}
where $q$ is the photon momentum.
Taking into account Lorentz invariance, current as well as parity conservation,
one arrives at the well-known decomposition of the matrix element in
Eq.~(\ref{Compton tensor}) in terms of  Tarrach's
amplitudes~\cite{Tarrach,Bernabeu:1976jq}. For our purposes, it is sufficient to consider
the process in the forward direction $p'=p$ and perform  spin-averaging in
Eq.~(\ref{Compton tensor}): 
\begin{equation}\label{Tensor_averaged}
T^{\mu\nu}(p,q)=\frac{1}{2}\sum_s T^{\mu\nu}(p,s;p,s;q).
\end{equation}
The tensor $T^{\mu\nu}(p,q)$  is related to the aforementioned invariant amplitudes
$T_1,\,T_2$ through the decomposition (see, e.g., Ref.~\cite{Gasser:2015dwa}):
\begin{equation}\label{Taveraged}
T^{\mu\nu}(p,q)= T_1(\nu,q^2)K_1^{\mu\nu} +T_2(\nu,q^2)K_2^{\mu\nu},
\end{equation}
where the kinematic structures $K_1^{\mu\nu},\, K_2^{\mu\nu}$ read
\begin{align}\label{eq:K}
K_1^{\mu\nu}&= q^\mu q^\nu-g^{\mu\nu}q^2,\nonumber\\
K_2^{\mu\nu}&=  \frac{1}{m^2}\Big\{(p^\mu q^\nu+p^\nu q^\mu)p\cdot q
-g^{\mu\nu}(p\cdot q)^2-p^\mu p^\nu q^2\Big\}.
\end{align}
Here, $\nu\equiv p\cdot q/m$. 

According to the asymptotic behavior of the structure functions at large values
of the parameter $\nu$,  the dispersion relation for the amplitude $T_1(\nu,q^2)$
requires one subtraction. It is usually performed at $\nu=0$, and hence the subtraction
function $S_1$ is defined as
\begin{equation}
S_1(q^2)=T_1(0,q^2)\,.
\end{equation}
The function $S_1(q^2)$ can be formally split in two parts:
\begin{equation}\label{S1}
S_1(q^2)=S_1^{\rm el}(q^2)+S_1^{\rm inel}(q^2)\,.
\end{equation}
The elastic term $S_1^{\rm el}(q^2)$ is associated with the one-nucleon exchange
in the $s$- and $u$-channels. The inelastic piece $S_1^{\rm inel}(q^2)$ is a regular
function of $q^2$.
We use the same definition of the elastic part as in Ref.~\cite{Agadjanov:2016cjc}:
\begin{eqnarray}\label{S1el}
S^{\rm el}_1(q^2)=-\frac{4m^2}{q^2(4m^2-q^2)}(G_E^2(q^2)-G_M^2(q^2))\,.
\end{eqnarray}
Little information is available on the inelastic part of the subtraction function $S_1^{\rm inel}(q^2)$.  
According to the low-energy theorem, its value at $q^2=0$ is given by
\begin{equation}\label{LET}
S_1^{\rm inel}(0)=-\frac{\kappa^2}{4m^2}-\frac{m}{\alpha}\, \beta_M\,.
\end{equation}
Here, $\beta_M$ denotes the magnetic polarizability of the nucleon,
$\alpha\simeq 1/137$ is the fine structure constant and $\kappa= F_2(0)$ is the anomalous
magnetic moment of the nucleon. At large values of $q^2$, the asymptotic behavior
of the subtraction function is
fixed by the operator product expansion (see, e.g., Ref.~\cite{Erben:2014hza,Hill:2016bjv}).
Otherwise, it is unknown in the intermediate kinematic region
$0< -q^2\lesssim$ 2 GeV$^2$, which is amenable to lattice simulations.

\subsection{External field configuration}
\label{sec:field}

In Ref.~\cite{Agadjanov:2016cjc},  it was proposed to place the nucleon in the
time-independent periodic magnetic field
\eq\label{Bperiodic}
{\bf B}=(0,0,B^3),\quad B^3=-eB\cos(\omega x^2)\,, 
\en	
where $B$ denotes the strength  of the field and the frequency $\omega$ takes  nonzero values. 
The components of the gauge field $A^\mu(x)$ are chosen as follows:
\eq\label{A-potential}
A^1=\frac{eB}{\omega}\sin(\omega x^2),\qquad A^0=A^2=A^3=0\,.
\en	
The magnetic flux is quantized in a finite box of size $L$:
\begin{equation}
\int_{-L/2}^{L/2}dx^1dx^2B^3(x^2)=6\pi N\,,
\end{equation}
As discussed in Ref.~\cite{Davoudi:2015cba}, this quantization condition can be
implemented on the lattice in two different ways. In the first scenario, the frequency
$\omega$ is constrained and no constraint is imposed on the magnetic field strength
$B$. In the second scenario, the situation is reversed. Hence, we have:
\begin{eqnarray}\label{QC1}
&\mbox{a)}&\qquad\omega=\frac{2\pi n}{L},\qquad n\in\mathbb{Z} \backslash \{0\},\quad  {\rm arbitrary} \,\,B,\\\label{QC2}
&\mbox{b)}&\qquad B=\frac{6\pi N}{eL^2}\frac{\omega L/2}{\sin (\omega L/2)},\qquad N\in \mathbb{Z} \backslash \{0\}, \quad  {\rm arbitrary} \,\,\omega\neq \frac{2\pi n }{L}\,. 
\end{eqnarray}
Only the first scenario was considered in Ref.~\cite{Agadjanov:2016cjc}.
In the present paper, we will exploit both quantization possibilities and demonstrate
that the obtained results are quite different. Obviously, the limit $\omega\to 0$ can be 
directly performed in the second setting only, where $\omega$ is a free parameter,
unrelated to the box size $L$.

\section{Non-relativistic framework}
\label{sec:NR}

\subsection{Method}

The framework, which is based on the use of the non-relativistic EFT, consists of two
steps. At the first stage, one matches the parameters of the non-relativistic Lagrangian
to the expression of the relativistic two-point function of the nucleon in an external
field. At the next step, one uses the resulting non-relativistic Hamiltonian to carry
out the calculation of the spectrum. The advantage of the method is  its transparency:
the calculations of the spectrum are done by using ordinary perturbation theory in
quantum mechanics. The setting is, however, not well suited for the calculation of the
finite-volume corrections, which are proportional to $\exp(-M_\pi L)$ where $M_\pi$
denotes the pion mass. Within this approach, these corrections should be included in the
couplings of the non-relativistic Lagrangian through the matching procedure.

Let us consider the two-point function of the nucleon, placed in the external field
$A^\mu(x)$. The path integral representation in  Minkowski space reads:
\begin{eqnarray}\label{2point}
\left\langle 0|T\Psi(x)\bar\Psi(y)|0\right\rangle _A=\frac{\displaystyle\int {\cal D}G{\cal D}q{\cal D}\bar q \,\Psi(x)\bar\Psi(y)e^{i\int d^4x({\cal L}+A_\mu(x)j^\mu(x) )}}{\displaystyle\int {\cal D}G{\cal D}q{\cal D}\bar q \,e^{i\int d^4x({\cal L}+A_\mu(x)j^\mu(x) )}},
\end{eqnarray}
where the integration over all possible gluon, quark and
antiquark field configurations is performed. Further,
$j_\mu(x)$ is the electromagnetic current, built of the quark fields, and $\Psi(x)$ denotes
the composite nucleon field operator in QCD. Expanding the right-hand-side of
Eq.~(\ref{2point}) up-to-and-including $O(A^2)$, one obtains
\begin{eqnarray}\label{eq:PsiPsi}
\langle 0|T\Psi(x)\bar\Psi(y)|0\rangle_A&=&\langle 0|T\Psi(x)\bar\Psi(y)|0\rangle_0
+\frac{i}{1!}\int d^4zA_\mu(z)\langle 0|T\Psi(x)\bar\Psi(y)j^\mu(z)|0\rangle_0
\nonumber\\[2mm]
&+&\frac{i^2}{2!}\int d^4z_1d^4z_2A_\mu(z_1)A_\nu(z_2)\langle 0|T\Psi(x)\bar\Psi(y)j^\mu(z_1)j^\nu(z_2)|0\rangle_0+\cdots\, ,
\end{eqnarray}
where the subscript
``0'' refers to the quantities evaluated in QCD without any external field, and
we have used the fact that $\left\langle 0|j^\mu(x)|0\right\rangle _0=0$. 
Note that the expansion in Eq.~(\ref{eq:PsiPsi})  is written down for connected matrix elements 
(the subscript ``conn'' is omitted everywhere for brevity).

In the above quantities, the nucleons are in general off the mass shell. Performing the Fourier transform in 
Eq.~(\ref{eq:PsiPsi}), amputating the external nucleon legs, and putting the
external nucleons on the mass shell,
we see that the nucleon electromagnetic vertex $\langle p',s'|j^\mu(0)|p,s\rangle$
emerges at order $A$. At order $A^2$, the Compton tensor, defined in Eq.~(\ref{Compton tensor}), is obtained
from the matrix element 
$\langle 0|T\Psi(x)\bar\Psi(y)j^\mu(z_1)j^\nu(z_2)|0\rangle_0 $.
 Note that, in general, the described procedure
  is equivalent to replacing the nucleon fields $\Psi(x)$ and $\bar\Psi(y)$ by the out- and ingoing nucleon states,
  $\langle p',s'|$ and $|p,s\rangle$ respectively. This fixes the overall normalization of
  the quantity we are considering below.
  
\subsection{Matching}

The first few terms of the effective Lagrangian, which describe
the interaction of the nucleon with an external electromagnetic field,
are given by:
\begin{equation}\label{L_eff}
  {\cal L}_{\rm eff}=\psi^\dagger\biggl(iD_t-m+\frac{{\bf D}^2}{2m}+\cdots\biggr)\psi
  -\frac{\mu}{e}\,\psi^\dagger \boldsymbol{\sigma}\cdot\mathbf{B}\psi + \frac{2\pi}{e^2}\,\psi^\dagger({\alpha}_E\mathbf{E}^2+ {\beta}_M\mathbf{B}^2)\psi +\cdots,
\end{equation}
where the ellipses denote higher order terms with derivatives, 
\begin{eqnarray}
D_t=\p_t-igA^0\,,\quad \m D=\nabla + ig \m A\, .
\end{eqnarray}
Here, $\psi(x)$ denotes the two-component nucleon field, $\mu$ is the magnetic
moment and $\alpha_E,\beta_M$ are the electric and magnetic polarizabilities, respectively.
The coupling constant $g$ takes the values $g=0,+1$ for the neutron and proton,
respectively, and ${\bf E},{\bf B}$ are the electric and magnetic fields (in order to
simplify the notations, we include the factor $e$ in the definition of the vector-potential
$A^\mu$). The NRQED
Lagrangian at order $m^{-4}$ is given, e.g., in Ref.~\cite {Hill:2012rh}.

If the properties of a nucleon {\em at rest} are calculated, it suffices to write down only
the first few terms in the Lagrangian. However, we are studying the nucleon in a periodic
field, with $\omega$ corresponding to the magnitude of the momentum transfer from the
field to the nucleon. Consequently, we have to retain {\em all} terms in the derivative
expansion of the Lagrangian given by Eq.~(\ref{L_eff}).
In this case, the exact
relativistic dispersion relation for the energy of the free nucleon with the
three-momentum $\m p$, $w(\m p)=\sqrt{m^2+\m p^2}$, is satisfied.

An important remark is in order. In the infinite volume, one is allowed to use  partial
integration in the Lagrangian. The same is true in a finite volume, if all fields 
(including the external electromagnetic field) are subject
to  periodic boundary conditions, since the surface terms vanish in this case. 
The above remark will be relevant, if the realization of the external field
is carried out according to the scenario b) from section~\ref{sec:field}: the
Lagrangians, which differ only by surface terms and lead to the same amplitudes
in the matching condition, might yield a different spectrum in the finite volume.
Bearing this in mind,
we must for instance
ensure that all terms of the Lagrangian that we write down are {\em explicitly}
gauge-invariant (not only up to the surface terms), otherwise, one is not guaranteed
that the resulting finite-volume spectrum is gauge-invariant. Note also that we did
not pay special attention to this issue in our previous
paper~\cite{Agadjanov:2016cjc}, where it was anyway not relevant, since only
the scenario a) was considered.

Taking the above issue into account, below we write down the explicit non-relativistic
Lagrangian, describing the nucleon in an external field up-to-and-including $O(A^2)$
in this field. Note also that we change the normalization of the fermion field by a factor
$(2W)^{1/2}$ with ${W}=\sqrt{m^2-\m D^2}$
as compared to Eq.~(\ref{L_eff}), in order to ensure that the free
one-nucleon states obey the relativistic normalization condition
(see, e.g., Refs. \cite{Colangelo:2006va,Gasser:2011ju}).
The Lagrangian takes the following form:
\eq\label{L_eff_A}
{\cal L}_{\rm eff}={\cal L}_0+{\cal L}_1+{\cal L}_2+\cdots\,.
\en
Here, 
\eq\label{eq:L0}
{\cal L}_{0}=\psi^\dagger(2{W})^{1/2}(iD_ t-{W})(2W)^{1/2}\psi\, ,
\en
\eq\label{eq:L1}
{\cal L}_1&=&\sum_{m,n=0}^\infty\,[\partial_{\mu_1}\cdots\partial_{\mu_n}E^j(x)]\,
[\psi_{s'}^\dagger(x)\stackrel{\leftrightarrow}{D^{i_1}}\cdots
\stackrel{\leftrightarrow}{D^{i_m}}
\Gamma_{E,s's}^{j,\mu_1\cdots \mu_n,\,i_1\dots i_m}\psi_{s}(x)]
\nonumber\\[2mm]
&+&\sum_{m,n=0}^\infty\,[\partial_{\mu_1}\cdots\partial_{\mu_n}B^j(x)]\,
[\psi_{s'}^\dagger(x)\stackrel{\leftrightarrow}{D^{i_1}}\cdots
\stackrel{\leftrightarrow}{D^{i_m}}
\Gamma_{B,s's}^{j,\mu_1\cdots \mu_n,\,i_1\dots i_m}\psi_{s}(x)]
\en
and
\eq\label{eq:L2}
{\cal L}_2&=&\sum_{m,n,k=0}^\infty\,[\partial_{\mu_1}\cdots\partial_{\mu_n}E^j(x)]\,
[\partial_{\nu_1}\cdots\partial_{\nu_k}E^l(x)]\,
[\psi_{s'}^\dagger(x)\stackrel{\leftrightarrow}{D^{i_1}}\cdots
\stackrel{\leftrightarrow}{D^{i_m}}
\Pi_{EE,s's}^{jl,\mu_1\cdots \mu_n,\,\nu_1\cdots\nu_k,\,i_1\dots i_m}\psi_{s}(x)]
\nonumber\\[2mm]
&+&\sum_{m,n,k=0}^\infty\,[\partial_{\mu_1}\cdots\partial_{\mu_n}E^j(x)]\,
[\partial_{\nu_1}\cdots\partial_{\nu_k}B^l(x)]\,
[\psi_{s'}^\dagger(x)\stackrel{\leftrightarrow}{D^{i_1}}\cdots
\stackrel{\leftrightarrow}{D^{i_m}}
\Pi_{EB,s's}^{jl,\mu_1\cdots \mu_n,\,\nu_1\cdots\nu_k,\,i_1\dots i_m}\psi_{s}(x)]
\nonumber\\[2mm]
&+&\sum_{m,n,k=0}^\infty\,[\partial_{\mu_1}\cdots\partial_{\mu_n}B^j(x)]\,
[\partial_{\nu_1}\cdots\partial_{\nu_k}B^l(x)]\,
[\psi_{s'}^\dagger(x)\stackrel{\leftrightarrow}{D^{i_1}}\cdots
\stackrel{\leftrightarrow}{D^{i_m}}
\Pi_{BB,s's}^{jl,\mu_1\cdots \mu_n,\,\nu_1\cdots\nu_k,\,i_1\dots i_m}\psi_{s}(x)]\, ,
\nonumber\\
\en
where $\Gamma_{E/B}$ and $\Pi_{EE/EB/BB}$ denote the pertinent combinations
of the effective couplings with the invariant tensors like
$g^{\mu\nu}$ or $\varepsilon^{\mu\nu\alpha\beta}$ and Pauli matrices for the spin
(in the following, for brevity,
we shall refer to $\Gamma_{E/B}$ and $\Pi_{EE/EB/BB}$ merely as to the effective
couplings). 
The Latin indices run from 1 to 3 (only space derivatives), whereas the Greek
indices run from 0 to 3.
The derivatives in the square brackets act only on the function within the brackets
and
\eq
\psi^\dagger \stackrel{\leftrightarrow}{D^i}\psi \equiv \psi^\dagger
(-\stackrel{\leftarrow}{\partial^i}+\stackrel{\rightarrow}{\partial^i}+2igA^i)\psi\, .
\en
Also, as a convention, 
the values $m,n,k=0$ correspond to no derivatives in Eqs.~(\ref{eq:L1},\ref{eq:L2}). 
Expanding explicitly in powers of the external field $A$ and
{\em using  partial integration} and the equations of motion,
one may rewrite the above Lagrangian in a simpler
form, already displayed in Ref.~\cite{Agadjanov:2016cjc}.
\eq
{\cal L}=\bar{\cal L}_0+\bar{\cal L}_1+\bar{\cal L}_2+\cdots\, ,
\en
where
\eq
\bar{\cal L}_{0}=\psi^\dagger2 w(i\p_ t-w)\psi,\quad\quad
w=\sqrt{m^2-\bs\nabla^2}\, ,
\en
whereas
\eq \label{L1L2}\nonumber
	\bar {\cal L}_1&=&\sum_{m,n=0}^\infty\,A_\mu(x)[\p_{i_1}\cdots \p_{i_n}\psi_{s'}^\dagger(x)]\Gamma_{s's}^{i_1\dots i_n,\,j_1\cdots j_m,\,\mu}[\p_{j_1}\cdots \p_{j_m}\psi_{s}(x)]\, ,\\[2mm]
	\bar{\cal L}_2&=&\sum_{l,m,n=0}^\infty\,A_\nu(x)[\p_{\mu_1}\cdots \p_{\mu_l}A_\mu(x)]\,[\p_{i_1}\cdots \p_{i_n}\psi_{s'}^\dagger(x)]\Pi_{s's}^{i_1\cdots i_n,\,j_1\cdots j_m,\,{\mu_1}\dots{\mu_l},\,\mu\nu}
	[\p_{j_1}\dots \p_{j_m}\psi_{s}(x)]\, .
        \nonumber\\
	\en
        Here, the effective couplings $\Gamma$ and $\Pi$ are the linear combinations of
        the couplings appearing in Eqs.~(\ref{eq:L0},\ref{eq:L1},\ref{eq:L2}).
        This form of the effective Lagrangian is better suited for carrying out the matching to the
        relativistic amplitudes. For example, the couplings $\Gamma$
 should be matched to the current
matrix element in Eq.~(\ref{current matrix}). Calculating the same vertex function
in the effective field theory with the Lagrangian $\bar {\cal L}_1$, we get
\begin{equation}\label{matching_A}
\sum_{m,n=0}^\infty\,(-i p')_{i_1}\dots (-i p')_{i_n}(i p)_{j_1}\dots (i p)_{j_m}
\Gamma_{s's}^{i_1\dots i_n,\,j_1\dots j_m,\,\mu}=\langle  p',s' |j^\mu(0)| p, s\rangle\, .
\end{equation}
This means that, expanding the nucleon form factor in a Taylor series in ${\bf p}$ and
${\bf p}'$,  one can determine all coefficients
$\Gamma_{s's}^{i_1\dots i_n,\,j_1\dots j_m,\,\mu}$. The matching at $O(A)$ is thus complete.

The matching at $O(A^2)$ proceeds along a similar pattern. The second-order term in
the expansion of the relativistic amplitude, given in Eq.~(\ref{eq:PsiPsi}), on the mass shell
can be written in the following form:
\begin{eqnarray}
M=\int \frac{d^4q}{(2\pi)^4}d^4z_1d^4z_2 e^{-iqz_1+i(p'-p+q)z_2}A_\mu(z_1)A_\nu(z_2)T^{\mu\nu}(p',s';p,s;q)\, ,
\end{eqnarray}
where $T^{\mu\nu}(p',s';p,s;q)$ is the Compton tensor defined in
Eq.~(\ref{Compton tensor}). On the other hand, in the non-relativistic theory,
there are two contributions at order $O(A^2)$: $M=M_1+M_2$.
The tree level contribution $M_1$ is given by the second order Lagrangian $\bar {\cal L}_2$:
\begin{eqnarray}
M_1&=&\sum_{l,m,n=0}^\infty\,\int \frac{d^4q}{(2\pi)^4}d^4z_1d^4z_2 e^{-iqz_1+i(p'-p+q)z_2}A_\mu(z_1)A_\nu(z_2)\nonumber\\[2mm]
&\times&(-ip')_{i_1}\dots (-ip')_{i_n}(ip)_{j_1}\dots (ip)_{j_m}(iq)_{\mu_1}\dots (iq)_{\mu_l}\Pi_{s's}^{i_1\dots i_n,\,j_1\dots j_m,\,{\mu_1}\dots{\mu_l},\,\mu\nu}\,.
\end{eqnarray}
The second iteration of the Lagrangian $\bar {\cal L}_1$ gives another term, $M_2$, with
\begin{eqnarray}
M_2=\int \frac{d^4q}{(2\pi)^4}d^4z_1d^4z_2 e^{-iqz_1+i(p'-p+q)z_2}A_\mu(z_1)A_\nu(z_2)U^{\mu\nu}(p',s';p,s;q),
\end{eqnarray}
where the tensor $U^{\mu\nu}(p',s';p,s;q)$ is given by the sum of the nucleon pole terms:
\eq\label{Compton tensor_eff}
  U^{\mu\nu}(p',s';p,s;q)&=&
  \dfrac{ \dsum\limits_\sigma\langle p',s'|j^\mu(0)|p'+q,\sigma\rangle
\langle p'+q,\sigma|j^\nu(0)|p,s\rangle}{4w({\bf p}'+{\bf q})(w({\bf p}'+{\bf q})-w({\bf p}')-q_0-i0)}\,
\nonumber\\[2mm]
  &+&   \dfrac{ \dsum\limits_\sigma\langle p',s'|j^\nu(0)|p-q,\sigma\rangle
\langle p-q,\sigma|j^\mu(0)|p,s\rangle}{4w({\bf p}-{\bf q})(w({\bf p}-{\bf q})-w({\bf p})+q_0-i0)}\, .
\en
Note that, in order to derive the above expression, the matching at $O(A)$ has been used.  

Finally, the matching condition at $O(A^2)$ reads:
\begin{eqnarray} \label{mA2}
&&\sum_{l,m,n=0}^\infty\,(-ip')_{i_1}\dots (-ip')_{i_n}(ip)_{j_1}\dots (ip)_{j_m}(iq)_{\mu_1}\dots (iq)_{\mu_l}\Pi_{s's}^{i_1\dots i_n,\,j_1\dots j_m,\,{\mu_1}\dots{\mu_l},\,\mu\nu}\nonumber\\[2mm]
&&=T^{\mu\nu}(p',s';p,s;q)-U^{\mu\nu}(p',s';p,s;q).
\end{eqnarray}
It is seen that the low-energy constants
$\Pi_{s's}^{i_1\dots i_n,\,j_1\dots j_m,\,{\mu_1}\dots{\mu_l},\,\mu\nu}$ are uniquely
determined by the nucleon pole-subtracted Compton scattering amplitude in QCD.

An important remark is in order. The aim of the matching is to determine the
couplings $\Gamma$ and $\Pi$, which encode the physics at short distances.
It can be carried out in the infinite volume, where no specific care about the
quantization of the magnetic flux should be taken. The latter will be, however,
important in the calculation of the energy shift.

\subsection{Perturbation theory for the energy levels}

In the previous section, an effort was made to match the relativistic and
non-relativistic theories at order $A^2$. In this section, we shall be rewarded for this effort,
using the resulting non-relativistic Lagrangian for the calculation of the energy
spectrum of the nucleon in an external field. Also, up to this moment, we have not
specified the external field. Here we assume that $A_\mu$ is the static field
described in section~\ref{sec:field} and the scenario a) is chosen.
Stationary energy levels exist in such background field configurations.

Consider the canonical Hamiltonian ${\cal H}$,
which is obtained from the non-relativistic Lagrangian ${\cal L}$. In order
to arrive at the non-relativistic normalization of states, used in quantum mechanics,
it is convenient to rescale
back the nucleon field, entering in this Hamiltonian,
as $\psi\to(2w)^{-1/2}\psi$.
Further, we define the quantum-mechanical Hamiltonian $H$, which is given by the
matrix element of  ${\cal H}$ between the free one-nucleon states.  $H$
is a differential operator, which acts on the nucleon wave function:
\begin{equation}
H=H_0+H_1+H_2+O(A^3)\, ,
\end{equation}
where
\begin{eqnarray}\label{Hamiltonian}
(H_0)_{s's}&=&w(\rnabla)\delta_{s's},\nonumber\\[2mm]
(H_1)_{s's}&=&-\frac{1}{\sqrt{2w(\lnabla)}} \sum_{m,n=0}^\infty\,\lp_{i_1}\dots \lp_{i_n}\,\Gamma_{s's}^{i_1\dots i_n,\,j_1\dots j_m,\,\mu}A_\mu(\m x)\rp_{j_1}\dots \rp_{j_m}\frac{1}{\sqrt{2w(\rnabla)}},\nonumber\\[2mm]
(H_2)_{s's}&=&-\frac{1}{\sqrt{2w(\lnabla)}} \sum_{l,m,n=0}^\infty\,\lp_{i_1}\dots \lp_{i_n}\,\Pi_{s's}^{i_1\dots i_n,\,j_1\dots j_m,\,{\mu_1}\dots{\mu_l},\,\mu\nu}[\rp_{\mu_1}\dots \rp_{\mu_l}A_\mu(\m x)]A_\nu(\m x)\nonumber\\[2mm]
&\times&\rp_{j_1}\dots \rp_{j_m}\frac{1}{\sqrt{2w(\rnabla)}}~.
\end{eqnarray}
Note that $H$ is a $2\times 2$ matrix in spin space.

The wave function of the nucleon in the external field obeys  the  Schr\"odinger equation:
 \begin{eqnarray}\label{Schr_eq}
H_{ss'}\psi_{n,s'}(\m x)=E\psi_{n,s}(\m x)\,,
\end{eqnarray}
where the $\psi_{n,s}(\m x)$ denote stationary solutions in a finite volume, satisfying  periodic boundary conditions.
The eigenfunctions $\psi^{(0)}_{n,s}(\m x)$ and the eigenvalues $w(\m k_n)$  of the unperturbed Hamiltonian $H_0$ satisfy the equation
\begin{eqnarray}
(H_0)_{ss'}\psi^{(0)}_{n,s'}(\m x)=w(\m k_n)\psi^{(0)}_{n,s}(\m x)\,.
\end{eqnarray}
The unperturbed spectrum has the form
\eq
w(\m k_n)=\sqrt{m^2+\m k_n^2}, \qquad
\m k_n=\frac{2\pi \bf n}{L}, \qquad \m n\in \mathbb{Z}^3\,,
\en
and the  normalized solutions are given by
\begin{equation}\label{wf_0}
\psi^{(0)}_{n,s}(\m x)\equiv\llangle{\bf x}|{\bf k}_n, s\rrangle=\frac{1}{L^{3/2}}e^{i\m k_n\m x}\chi_s\,,\quad \llangle {\bf k}_{m},s'|{\bf k}_n,s\rrangle 
=\delta_{{\bf m}{\bf n}}\delta_{s's}\, .
\end{equation}
Here, we have introduced a double-bracket
notation that is different from the relativistic case.
Namely, $|{\m k_n, s}\rrangle$ denotes the state vector in the non-relativistic theory,
which corresponds to the unperturbed solution.

Next, we apply  perturbation theory in order to calculate the shift of the
ground state in the external field. By doing this, we implicitly assume that the structure
of the spectrum is not changed by the background field which is sufficiently small.
Below, we shall put this condition under scrutiny.

Let us start from the ground state energy shift at order $A$. The unperturbed spectrum is
degenerate (the same energy for both spin projections), so the perturbation
theory for the degenerate levels should be applied. As is well known (see, e.g.,
Ref.~\cite{Landau}), the first-order energy shift is the solution of the secular equation:
\begin{eqnarray}\label{sec_eq}
\det(V_{s's}-\delta E^{(1)}\delta_{s's})=0\,,
\end{eqnarray}
where $V_{s's}= \llangle{\m 0,s'}|H_1|{\m 0, s}\rrangle$.

The matrix element of the operator $H_1$ is given by
\begin{eqnarray}\label{H1integral}
  \llangle \m p',s' |H_1 |\m p, s\rrangle &=&-\frac{1}{L^3}\int_{-L/2}^{L/2}d^3\m x\, e^{-i\m p'\m x} \frac{1}{\sqrt{2w(\lnabla)}}
  \nonumber\\[2mm]
&\times&\sum_{m,n=0}^\infty\,\lp_{i_1}\dots \lp_{i_n}\,\Gamma_{s's}^{i_1\dots i_n,\,j_1\dots j_m,\,\mu}A_\mu(\m x)\rp_{j_1}\dots \rp_{j_m}\frac{1}{\sqrt{2w(\rnabla)}}e^{i\m p\m x}\,.
\nonumber\\[2mm]
                                          &=&-\frac{\tilde{A}_\mu(\m p-\m p')}{L^3\sqrt{4w(\m p')w(\m p)}}\sum_{m,n=0}^\infty\,(-i\m p')_{i_1}\dots (-i\m p')_{i_n}(i\m p)_{j_1}\dots (i\m p)_{j_m}\Gamma_{s's}^{i_1\dots i_n,\,j_1\dots j_m,\,\mu}\, .
 \nonumber\\                                             
\end{eqnarray}
Here, $\tilde{A}_\mu(\m q)$ denotes the Fourier transform of the field $A_\mu(\m x,0)$
\begin{equation}\label{Ap}
\tilde{A}_\mu(\m q)=\int_{-L/2}^{L/2}d^3\m x\,e^{i\m q \m x}A_\mu(\m x,0),
\end{equation}
which, for the field configuration described in Eq.~(\ref{A-potential}), gives
\begin{eqnarray}\label{Atilde}
  \tilde{A}^1(\m q )=\frac{eB}{2i\omega}L^3[\delta_{\m q,-\boldsymbol \omega}-\delta_{\m q,\boldsymbol\omega}],\quad \tilde{A}^0=\tilde{A}^2=\tilde{A}^3=0,\quad
  \mb\omega=(0,\omega,0)\neq 0\, .
\end{eqnarray}
The sum in Eq.~(\ref{H1integral}) has precisely the same form as in the matching condition at $O(A)$, Eq.~(\ref{matching_A}).
Accordingly, the matrix element of the  operator $H_1$ takes the form:
\begin{eqnarray}\label{H1matrix element}
\llangle\m  p',s' |H_1 |\m p, s\rrangle =-\frac{\langle p',s' |j^1(0)|p, s\rangle }{\sqrt{4w(\m p')w(\m p)}}\frac{1}{L^3}\tilde{A}^1(\m p-\m p')\, .
\end{eqnarray}
Setting $\m p=\m p'=0$ in Eq. (\ref{H1matrix element}) and taking into account that
$\omega\neq 0$, it is seen that the
matrix elements $V_{ss'}$ vanish:
\begin{eqnarray}
V_{ss'}=0\,.
\end{eqnarray}
Thus, there is no first-order correction to the  energy shift:
\begin{equation}\label{dEfirst}
\delta E^{(1)}=0\,.
\end{equation}
As expected, this result follows from the three-momentum conservation at the vertex of the three-point function. We again stress that it holds only for $\omega\neq 0$.
Note also that, since the off-diagonal matrix elements vanish as well, the correct wave functions at this order are still given by Eq.~(\ref{wf_0}).

The second-order contribution to the energy shift can be found again from the secular equation, which differs from Eq. (\ref{sec_eq}) by the replacement
\begin{eqnarray}
V_{s's}\rightarrow \sum_{\m k_n\neq\,\m 0}\sum_\sigma\,\frac{\llangle {\m 0,s'}|H_1|{\m k_n, \sigma}\rrangle \llangle {\m k_n,\sigma}|H_1|{\m 0, s}\rrangle }{w(\m 0)-w(\m k_n)}+\llangle \m 0,s' |H_2 |\m 0, s\rrangle\,.
\end{eqnarray}
The first term emerges
from the second iteration of $H_1$ and another one is the matrix element of $H_2$. The spin-averaged energy correction at $O(B^2)$ 
consists of two pieces:
\begin{eqnarray}
\frac{1}{2}\,\sum_s\delta E_s^{(2)}=\frac{1}{2}\,\sum_s(\delta E_s'+\delta E_s{''})\, ,
\end{eqnarray}
where 
\eq
\delta E_s'&=&\sum_{\m k_n\neq\,\m 0}\sum_\sigma\,\frac{\llangle {\m 0,s}|H_1|{\m k_n, \sigma}\rrangle \llangle {\m k_n,\sigma}|H_1|{\m 0, s}\rrangle }{w(\m 0)-w(\m k_n)},
\nonumber\\[2mm]
\delta E_s''&=&\llangle {\bf 0},s|H_2|{\bf 0},s\rrangle\, .
\en
The first term is evaluated by using Eqs. (\ref{Atilde},\ref{H1matrix element}).
Taking into account the fact that $w({\bf 0})=m$, we obtain:
\begin{eqnarray}\label{eq:prime}
\delta E_s'=\frac{1}{4m}\left( \frac{eB}{\omega}\right)^2\sum_{\m k_n\neq\m 0}\sum_{\sigma}\,\frac{\langle \hat p,s |j^1(0)| k_n, \sigma\rangle\langle k_n, \sigma|j^1(0)|\hat p, s\rangle}{4w(\m k_n)(m-w(\m k_n))}\left[\delta_{\m k_n,-\boldsymbol \omega}+\delta_{\m k_n,\boldsymbol\omega}\right]\, ,
\end{eqnarray}
where $\hat p=(m,{\bf 0})$.
Performing the summation over $\m k_n$, we get:
\eq
\delta E_s'&=&\frac{(eB)^2}{8m\omega^2}
[F(\bs \omega)+F(-\bs\omega)]\, ,
\en
where the quantity $F(\bs \omega)$ reads
\eq
F(\bs \omega)=\sum_{\sigma}\frac{\langle \hat p,s |j^1(0)|\hat p+\hat q, \sigma\rangle\langle \hat p+\hat q, \sigma|j^1(0)|\hat p, s\rangle}{2w(\bs \omega)(m-w(\bs \omega))}\,,\qquad \hat q=(0,{\mb \omega})\, .
\en
For the second piece we need to evaluate the matrix element of the operator $H_2$:
\begin{eqnarray}\label{H2integral}
\llangle \m p',s' |H_2 |\m p, s\rrangle &=&-\frac{1}{L^3}\int_{-L/2}^{L/2}d^3\m x\, e^{-i\m p'\m x} \frac{1}{\sqrt{2w(\lnabla)}} \times\\\nonumber
&\times&\sum_{l,m,n=0}^\infty\,\lp_{i_1}\dots \lp_{i_n}\,\Pi_{s's}^{i_1\dots i_n,\,j_1\dots j_m,\,{\mu_1}\dots{\mu_l},\,\mu\nu}[\p_{\mu_1}\dots \p_{\mu_l}A_\mu(\m x)]A_\nu(\m x)\times\\
&\times&\rp_{j_1}\dots \rp_{j_m}\frac{1}{\sqrt{2w(\rnabla)}}e^{i\m p\m x}\,.
\end{eqnarray}
The integration leads to the expression
\begin{eqnarray}\label{H2matrix element}
\llangle \m p',s' |H_2 |\m p, s\rrangle&=&-\frac{1}{L^3\sqrt{4w(\m p')w(\m p)}} \\\nonumber
&\times&
\sum_{l,m,n=0}^\infty\,(-i\m p')_{i_1}\dots (-i\m p')_{i_n}(i\m p)_{j_1}\dots (i\m p)_{j_m}\Pi_{s's}^{i_1\dots i_n,\,j_1\dots j_m,\,{\mu_1}\dots{\mu_l},\,\mu\nu}\,
I_{\mu_1\dots\mu_l,\mu\nu}\, ,
\end{eqnarray}
where the integral $I_{\mu_1\dots\mu_l,\mu\nu}$ reads
\begin{equation}
I_{\mu_1\dots\mu_l,\mu\nu}=\int_{-L/2}^{L/2}d^3\m x\,e^{i{\bf q}{\bf x}}[\p_{\mu_1}\dots \p_{\mu_l}A_\mu(\m x)]A_\nu(\m x)\,.
\end{equation}
If ${\bf q}=0$, this integral has a non-zero value for $\mu_1=\dots=\mu_l=2$, $\mu=\nu=1$ and for even $l$,
\begin{equation}
I_{2\dots 2,11}=\frac{L^3}{2}\left( \frac{eB}{\omega}\right)^2(i\omega)^l, \qquad l=0,2,\dots\,.
\end{equation}
Inserting this expression into Eq. (\ref{H2matrix element}), one gets 
\begin{eqnarray}
\llangle\m p,s |H_2 |\m p, s\rrangle&=&-\frac{1}{4w({\bf p})}\left(\frac{eB}{\omega}\right)^2 \\\nonumber
&\times&
\sum_{l,m,n=0}^\infty\,(-i\m p)_{i_1}\dots (-i\m p)_{i_n}(i\m p)_{j_1}\dots (i\m p)_{j_m}\underbrace{(i\mb\omega)\dots (i\mb \omega)}_\textrm{$l$ copies}\Pi_{ss}^{i_1\dots i_n,\,j_1\dots j_m,\,{\mu_1}\dots{\mu_l},\,11}
\,.
\end{eqnarray}
Further, using the matching condition at $O(A^2)$ in Eq. (\ref{mA2}),
and setting  $\m p=0$, we see that the expression for the second energy correction $\delta E_s''$ takes a compact form:
\begin{eqnarray}
\delta E_s''=-\frac{1}{4m}\left(\frac{eB}{\omega}\right)^2\left[T^{11}(0,s;0,s;\hat q)-U^{11}(0,s;0,s;\hat q)\right]
\,,
\end{eqnarray}
where $\hat q=(0,\mb \omega)$.

It remains to put all pieces together. Noting that the quantity $U^{11}(0,s;0,s;\hat q)$,
defined by Eq.~(\ref{Compton tensor_eff}), is exactly equal to
$-\frac{1}{2}\,[F(\bs \omega)+F(-\bs \omega)]$, we finally arrive at the expression
of the spin-averaged energy shift of the ground state, derived first in Ref.~\cite{Agadjanov:2016cjc}:
\begin{equation}\label{finalresult}
\delta E=-\frac{1}{4m}\left( \frac{eB}{\omega}\right)^2\frac{1}{2}\sum_{s}\, T^{11}(0,s;0,s;\hat q)+O(B^3)=\frac{(eB)^2}{4m}\,T_1(0,-\omega^2)+O(B^3)
\, ,
\end{equation}

\subsection{The zero-frequency limit}\label{Zero_freq}

The equation (\ref{finalresult}) does not posses a smooth zero-frequency limit. This is seen from the fact that, e.g., the quantity $T_1(0,-\omega^2)$ includes 
the elastic contribution
that diverges in this limit as $1/\omega^2$. On the other hand, the result for the energy
shift in the constant field is well known: it is finite and is proportional to the pole-subtracted
part of the forward Compton amplitude. In this section, we shall discuss this apparent
contradiction.

Let us start with the energy shift at $O(A)$.
As the frequency of the magnetic field tends to zero, the field approaches 
a constant value and the first-order correction to the energy shift 
does not vanish anymore. It is immediately seen that approaching smoothly
the limit $\omega\to 0$ is not possible in the scenario a), where
$\omega$ is quantized, according to Eq.~(\ref{QC1}) so that $\omega$ is either
zero from the beginning or not. If $\omega\neq 0$, then the energy shift of
the ground state, caused by the perturbation Hamiltonian in 
Eq.~(\ref{H1matrix element}), is strictly zero. Consequently, one has to 
turn to scenario b). Here, one can immediately visualize the problem with
the non-vanishing surface terms, which were mentioned above. For example, 
the matrix element of the current, entering Eq.~(\ref{H1matrix element}),
has the following representation:
\eq\label{eq:vertex}
\langle p',s'|j^k(0)|p,s\rangle&=&\delta_{s's}(a_1(p'+p)^k+a_2q^k)
\nonumber\\[2mm]
&+&i\varepsilon^{kim}(a_3\sigma^m_{s's}q^i+a_4\sigma^m_{s's}(p'+p)^i
+a_5\sigma^l_{s's}(p'+p)^lq^i(p'+p)^m)\, ,
\en
where
\eq\label{eq:ai}
a_1&=&N\biggl((w({\bf p}')+w({\bf p})+2m)F_1(q^2)
-\frac{{\bf q}^2}{2m}\,F_2(q^2)\biggr)\, ,
\nonumber\\[2mm]
a_2&=&N(w({\bf p}')-w({\bf p}))\biggl(-F_1(q^2)
+\frac{w({\bf p}')+w({\bf p})}{2m}\,F_2(q^2)\biggr)\, ,
\nonumber\\[2mm]
a_3&=&-N\biggl((w({\bf p}')+w({\bf p})+2m)F_1(q^2)
+\frac{N^{-2}/2+2{\bf p}'{\bf p}}{2m}\,F_2(q^2)\biggr)\, ,
\nonumber\\[2mm]
a_4&=&N(w({\bf p}')-w({\bf p}))F_1(q^2)\, ,
\nonumber\\[2mm]
a_5&=&\frac{1}{2m}\,NF_2(q^2)\, ,
\en
and
\eq
N=\frac{1}{2\sqrt{(w({\bf p}')+m)}\sqrt{(w({\bf p})+m)}}\, .
\en
In the matching procedure, which is carried out in the infinite volume, we
always have ${\bf q}={\bf p}'-{\bf p}$ and no ambiguity arises. The same is true,
if the electromagnetic field potential obeys periodic boundary conditions, 
scenario a): in this case, the Dirac delta function, corresponding to the 
three-momentum conservation, is replaced by the Kronecker delta-symbol.
However, in case of a generic $\omega$, an ambiguity arises in the 
calculation of the energy shift, since the integration over the three-space does not
lead to the Kronecker delta and the three-momenta are no longer conserved.
In order to visualize the problem, consider, for instance, the case
of the proton where $F_1(0)=1$ and $F_2(0)=\kappa$ is the anomalous magnetic
moment. It suffices to retain
only the terms which are linear in the three-momenta in the expression of the matrix element
of the Hamiltonian $H_1$, given by Eq.~(\ref{H1matrix element}):
\eq\label{eq:H1}
&&\llangle {\bf p}',s'|H_1|{\bf p},s\rrangle=-\frac{1}{2mL^3}\,
\biggl(\delta_{s's}(p'+p)^k-i(1+\kappa)\varepsilon^{kim}q^i\sigma^m_{s's}\biggr)
\tilde A^k({\bf q})+\cdots
\\[2mm]
&=&\frac{eB}{4\omega mL}\biggl(\delta_{s's}(p'+p)^1-i(1+\kappa)q\sigma^3_{s's}\biggr)\biggl\{\frac{e^{i(\omega+q)\frac{L}{2}}-e^{-i(\omega+q)\frac{L}{2}}}{\omega+q}
+\frac{e^{-i(\omega-q)\frac{L}{2}}-e^{i(\omega-q)\frac{L}{2}}}{\omega-q}\biggr\}+\cdots\, ,
\nonumber
\en
where ${\bf q}=(0,q,0)$.
Assuming now ${\bf p}'={\bf p}={\bf 0}$ in this expression and then letting $\omega\to 0$
leads to the vanishing matrix element for any non-zero $\omega$. On the other
hand, using {\em partial integration,} one obtains
$\varepsilon^{kim}q^i\tilde A^k({\bf q})=-iB^m({\bf q})$, where $B$ is the magnetic field,
and the matrix element of the Hamiltonian takes the form:
\eq\label{eq:H1B}
\llangle {\bf p}',s'|H_1|{\bf p},s\rrangle&=&-\frac{1}{2mL^3}\,
\delta_{s's}(p'+p)^k\tilde A^k({\bf q})
+\frac{1+\kappa}{2mL^3}\,\sigma^k_{s's}\tilde B^k({\bf q})
\nonumber\\[2mm]
&=&
\frac{eB}{4\omega mL}\delta_{s's}(p'+p)^1
\biggl\{\frac{e^{i(\omega+q)\frac{L}{2}}-e^{-i(\omega+q)\frac{L}{2}}}{\omega+q}
+\frac{e^{-i(\omega-q)\frac{L}{2}}-e^{i(\omega-q)\frac{L}{2}}}{\omega-q}\biggr\}
\nonumber\\[2mm]
&+&\frac{i(1+\kappa)eB}{4mL}\,\sigma^3_{s's}
\biggl\{\frac{e^{i(\omega+q)\frac{L}{2}}-e^{-i(\omega+q)\frac{L}{2}}}{\omega+q}
-\frac{e^{-i(\omega-q)\frac{L}{2}}-e^{i(\omega-q)\frac{L}{2}}}{\omega-q}\biggr\}+\cdots\, .
\en
This expression does not vanish anymore and, in the limit $\omega\to 0$, yields the well-known
result for the first-order energy level splitting in the constant magnetic field.
The reason for this inequivalence is immediately seen: in case of an external field,
which does not obey periodic boundary conditions,
the 3-momentum conservation is not guaranteed, and the equality ${\bf q}={\boldsymbol\omega}$
does not hold anymore. At threshold, the vector ${\bf q}$ vanishes, but the vector potential
contains the factor $1/\omega$ and the result depends on the way the limit is performed. Indeed, since $q$ is always quantized, 
the difference between Eqs.~(\ref{eq:H1}) and (\ref{eq:H1B})
is proportional to
\eq
h(\omega)=\frac{e^{i\omega\frac{L}{2}}-e^{-i\omega\frac{L}{2}}}{\omega}\, .
\en
The quantity $h(\omega)=0$ if $\omega=({2\pi}/{L})\,n$ is quantized.
On the other hand, $h(\omega)\to iL$ for a fixed $L$ and $\omega\to 0$.

It can be also seen that the above ambiguity disappears, if an explicitly gauge-invariant Lagrangian, defined in Eqs.~(\ref{L_eff_A},\ref{eq:L0},\ref{eq:L1}) and 
(\ref{eq:L2}), is used from the beginning. The equation (\ref{eq:H1B}), which
leads to the correct result in the zero-frequency limit,
is directly obtained by using the gauge-invariant Lagrangian without performing
 the partial integration.

The situation is pretty much the same in case of the second-order energy shift.
Below, for simplicity, we shall consider the case of the neutron only.
Further, we shall stick to the particular field configuration, described in section~\ref{sec:field}. The nucleon pole contribution to the energy shift
(the analogue to Eq.~(\ref{eq:prime})) in case of the external field that does not obey
periodic boundary conditions, is given by
\eq\label{eq:wprime}
\delta E_s'=\frac{(eB)^2}{4m}
\sum_{\m k_n\neq\m 0}\sum_{\sigma}\,
\frac{\langle \hat p,s |j^1(0)| k_n, \sigma\rangle
\langle k_n, \sigma|j^1(0)|\hat p, s\rangle}{4w(\m k_n)(m-w(\m k_n))}\, 
f^2({\bf k}_n,{\boldsymbol\omega})\, ,
\en
where
\eq
f({\bf q},{\boldsymbol\omega})
=\frac{1}{\omega L^3}\,\int_{-L/2}^{L/2}d^3{\bf x}e^{i{\bf q}{\bf x}}
(e^{i{\boldsymbol\omega}{\bf x}}-e^{-i{\boldsymbol\omega}{\bf x}})\, .
\en
It is seen that, in case of the a periodic field, this factor does not reduce
to the Kronecker delta-symbol, corresponding to the conservation of the total
three-momentum. In the limit $\omega\to 0$ the above expression simplifies
considerably, and we have
\eq
f^2({\bf k}_n,{\boldsymbol\omega})
\to\frac{4}{{\bf k}_n^2}\,
\delta_{{\bf k}_n^\perp,{\bf 0}}\, ,
\en
where ${\bf k}_n^\perp$ denotes the components of the vector ${\bf k}_n$,
perpendicular to the vector ${\boldsymbol\omega}$.

The expression for the matrix elements, entering Eq.~(\ref{eq:wprime}), can be read off from Eqs.~(\ref{eq:vertex}) and (\ref{eq:ai}). First of all, because the three-momentum,
perpendicular to the direction of $\mb \omega$ is conserved, only the terms that
contain $a_3$ and $a_4$, can potentially contribute. 
Further, comparing Eq.~(\ref{eq:H1}) and Eq.~(\ref{eq:H1B}), it is clear that
using the gauge-invariant Lagrangian (\ref{eq:L1}) in our case boils down to the following
heuristic prescription: write down the vertices in terms of two linearly independent
vectors ${\bf p}'+{\bf p}$ and
${\bf q}={\bf p}'-{\bf p}$, and replace everywhere ${\bf q}$ through $\mb\omega$ as
if the three-momentum was conserved.
Now, one can ensure that the contributions
from both terms, containing either $a_3$ or $a_4$, 
vanish in the limit $\omega\to 0$. Indeed, as seen from Eq.~(\ref{eq:vertex}),
the term with $a_3$ contains ${\bf q}$, which is eventually
replaced by $\mb\omega$ and the limit $\omega\to 0$ is performed afterwards.
Further, using Eq.~(\ref{eq:ai}), one sees that $a_4$ is proportional to $w({\bf p}')-w({\bf p})$ or, equivalently, to ${\bf q}({\bf p}'+{\bf p})$. This expression also vanishes, when
${\bf q}$ is replaced by $\mb\omega$ and the limit $\omega\to 0$ is performed
(we remind the reader that the Fourier transform $A^1({\bf q})$ stays finite for
a non-zero ${\bf q}$ and $\omega\to 0$).
Hence, the entire pole
term does not contribute to the energy shift in the limit $\omega\to 0$, and
the latter is given solely by the contact contribution:
\eq
\delta E''_s=\llangle{\bf 0},s|H_2|{\bf 0},s\rrangle
=-\frac{1}{2mL^3}\,\Pi_{BB,ss}^{33}\int_{-L/2}^{L/2}d^3{\bf x}B^2({\bf x})\to
-\frac{(eB)^2}{2m}\,\Pi_{BB,ss}^{33}\, .
\en
Only a single coupling $\Pi_{BB,ss}^{33}$ contributes in this limit, since
the derivative terms give vanishing contributions. Comparing
the first term in the expansion of Eq.~(\ref{eq:L2}) to Eq.~(\ref{L_eff}) and taking
into account the different normalization of the nucleon field in these two Lagrangians,
one immediately sees that $\Pi_{BB,ss}^{33}$ is given by the magnetic polarizability
\eq
\Pi_{BB,ss}^{33}=\frac{m\beta_M}{\alpha}
\en
and, thus, the standard formula for the spin-averaged energy shift $\delta E=-2\pi\beta_MB^2$ is 
reproduced in the limit $\omega\to 0$ (note that $\delta E''_s$, given by the above expression, does not depend on the spin orientation).

To summarize this part, we note that, in order to perform a smooth
zero-frequency limit, one has to use the realization b) of the external
field on the lattice, in which the frequency $\omega$ is not quantized.
Using this realization for a finite $\omega$, however, is not very convenient.
Apart from the subtleties, arising in the treatment of the surface terms,
the final expression for the energy shift 
is rather complicated and simplifies only in the limit $\omega\to 0$.
For this reason, in the following we stick to the scenario a).

\subsection{Landau levels}

Here we consider, how the Landau levels emerge
from the periodic potential in the zero frequency limit $\omega\to 0$ . Further, we
give an estimate for the maximum value of the field strength $B$, for which our 
method still works (note that a crude estimate was provided already in our 
first paper~\cite{Agadjanov:2016cjc}). In order to simplify the discussion, we
merely discard the whole string of non-minimal couplings of the (charged)
nucleon to the external field, since they only give  corrections to the
Landau levels (in the zero frequency limit). 

We look for a stationary solution of the Dirac equation (see, e.g., \cite{IZ}):
\begin{eqnarray}
\left(i\slashed{\partial}+g\slashed{A}-m-\frac{\kappa}{4m}\sigma_{\mu\nu} F^{\mu\nu}\right) \psi(x)=0\,,
\end{eqnarray}
where $F^{\mu\nu}=\p^\mu A^\nu-\p^\nu A^\mu$ denotes the electromagnetic field strength tensor and $g=+1,0$ for the proton and the neutron, respectively.
Writing the wave function $\psi$ as
\begin{equation}
\psi(\m x,t)=e^{-iEt}
\begin{pmatrix}F(\m x)
\\ 
G(\m x)
\end{pmatrix}\,,
\end{equation}
we obtain:
\begin{eqnarray}\label{F}
\left( E-m+\frac{\kappa}{2m}\sigma\cdot \m B\right) F=\mb\sigma({\m p}+g\m A)G\\[2mm]\label{G}
\left( E+m-\frac{\kappa}{2m}\sigma\cdot \m B\right) G=\mb\sigma({\m p}+g\m A)F\,,
\end{eqnarray}
with ${\m p}=-i\mb\nabla$.
Further, it is convenient to consider the non-relativistic limit, 
in which the mass $m$ is the largest term on the l.h.s of (\ref{G}). 
Assuming that $eB\ll m^2$, 
the function $G$ can be easily expressed from Eq. (\ref{G})
\begin{eqnarray}
G\approx \frac{1}{2m}\mb\sigma({\m p}+g\m A)F\,,
\end{eqnarray}
and hence one gets an equation for $F(\m x)$:
\begin{eqnarray}\label{Deq_p}
[({\m p}+g\m A)^2-(g+\kappa)\mb\sigma\cdot \m B]F(\m x)=(E^2-m^2)F(\m x)\,.
\end{eqnarray}
For the field configuration given in Eq. (\ref{Bperiodic}), the solution $F(\m x)$ can be searched by using the ansatz
\begin{equation}
F(\m x)=e^{i(p_1x_1+p_3x_3)}\,f(x_2)\,,
\end{equation}
where $p_1,\,p_3$ are  the conserved components of the  three-momentum, and $f$ is also an eigenvector of the $\sigma_3$ matrix:
\begin{equation}
\sigma_3f=\alpha f,\qquad \alpha=\pm 1\,.
\end{equation}
Using now the periodic field configuration from Eqs.~(\ref{Bperiodic}) and
(\ref{A-potential}),  Eq. (\ref{Deq_p}) takes the form 
\begin{equation}\label{eq:starting}
\left[-\frac{d^2}{dx_2^2}+\frac{e^2g^2B^2}{\omega^2}\sin^2(\omega x_2) +\alpha(g+\kappa) eB\cos(\omega x_2) \right]f(x_2)=(E^2-m^2)f(x_2)\,,\quad \alpha=\pm 1\,.
\end{equation}
where we have set $p_1=p_3=0$ to focus on the ground state.

Let us first consider the case of the proton with $g=1$.
Introducing a new variable $z=\omega x_2/2$, this equation
can be brought to the standard form of the Whittaker-Hill equation:
\begin{equation}\label{WH}
\left[\frac{d^2}{dz^2}+a+2p\cos(4z)-2q\cos(2z) \right]f(z)=0\,,
\end{equation}
where 
\begin{equation}\label{aqq'}
a=\frac{4}{\omega^2}\left( E^2-m^2-\frac{e^2B^2}{2\omega^2}\right)\, ,
\quad q=\alpha \frac{2(1+\kappa) eB}{\omega^2}\, ,
\quad p=\frac{e^2B^2}{\omega^4}\,.
\end{equation}
According to Floquet's theorem (see, e.g., Ref.~\cite{AS}), Eq. (\ref{WH}) has  solutions with the pseudoperiodic property
\begin{equation}
f(z+\pi)=e^{i\nu \pi}f(z)\,,
\end{equation}
where $\nu$ denotes the characteristic exponent. As is well known, solutions are bounded only for certain values of $a$ (the parameters $p,q$ are fixed), which form the band structure. Only in this case, $\nu$ has a vanishing imaginary part.

\begin{figure}
	\centering
	\includegraphics[width=0.8\linewidth]{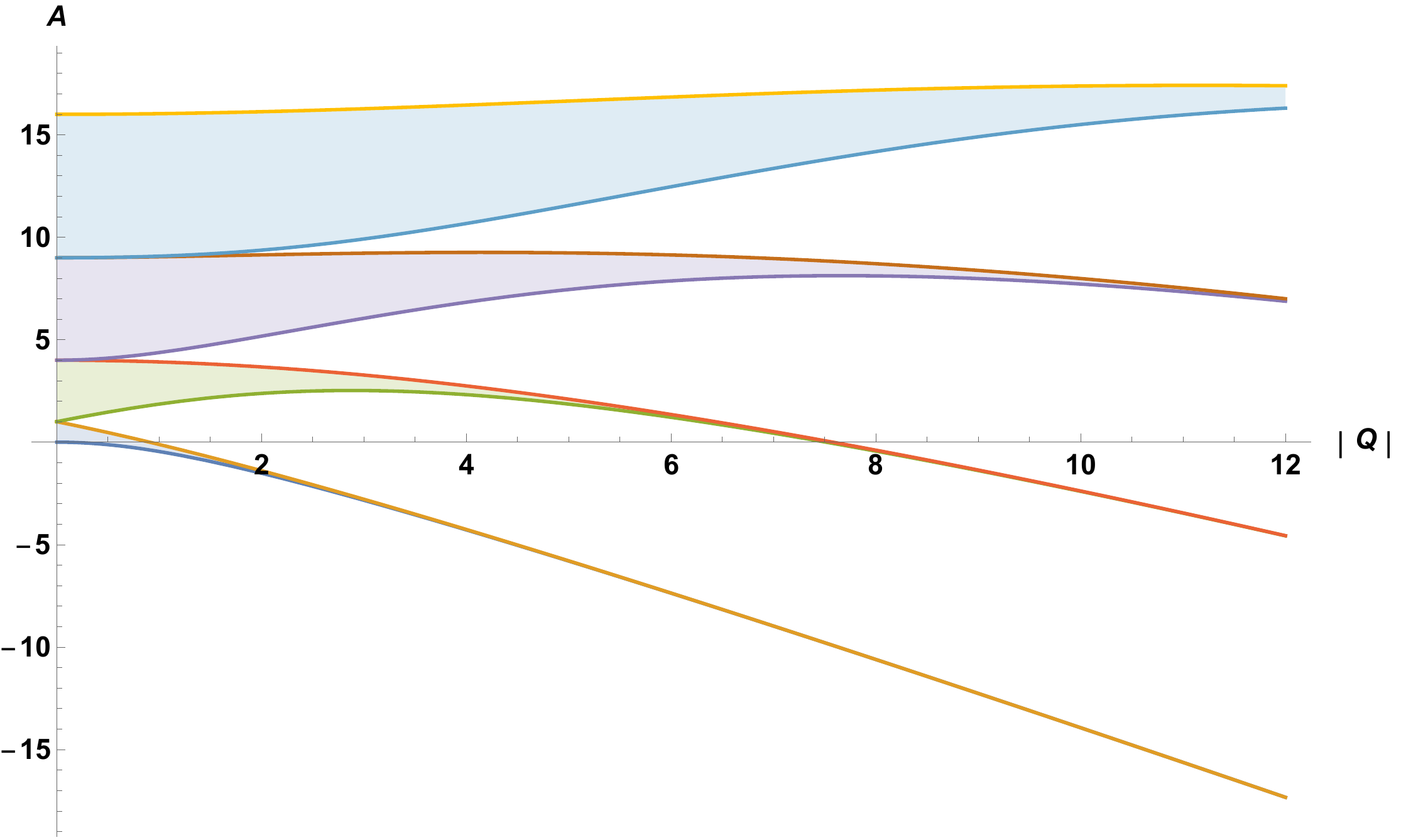}
	\caption{Stability chart of the Mathieu equation. In the colored regions (bands), the solutions are bounded. Landau levels emerge from the band spectrum as $|Q|\rightarrow\infty$.}
	\label{Mathieu}
\end{figure}

To see how the Landau levels emerge in the limit $\omega\rightarrow 0$, it is useful
to go back to Eq.~(\ref{eq:starting}). It is clear that, in the limit $\omega\to 0$,
the cosine in the last term on the left-hand-side can be replaced by unity.
Then, Eq. (\ref{WH}) takes the form of the Mathieu equation:
\begin{equation}
\left[\frac{d^2}{dz'^2}+A-2Q\cos(2z')\right]f(z')=0\,,
\end{equation}
where $z'=2z$ and $A=(a-2q)/4$, $Q=-p/4$. In Fig. \ref{Mathieu} we show the stability chart for this equation. In particular, for the colored regions in the $A-Q$ plane, the solutions are bounded. The band structure is clearly seen for $Q\neq 0$. As $Q\rightarrow\infty$, each band
smoothly transforms into a Landau level. We note that the stability chart for Eq.~(\ref{WH}) will be slightly different, but the picture is very similar, in particular, in the large $Q$ region. This qualitative result can be explicitly verified by using the asymptotic expansion of the eigenvalues $A_n$ ($n=0,1,\dots$) for large $Q$ (see, e.g., \cite{AS}):
\begin{eqnarray}
A_n=-2|Q|+2\sqrt{|Q|}(2n+1)\,,
\end{eqnarray}
or,
\begin{eqnarray}
E^2=m^2+|eB|(2n+1)+\alpha(1+\kappa)eB\,,\quad \alpha=\pm 1\,.
\end{eqnarray}

The same conclusion can be drawn in a finite volume. In this case, the solutions
in a band (the Bloch wave functions), do not obey, in general,  periodic boundary conditions. The requirement of
periodicity, $f(z+\pi)=f(z)$, picks out one level from the band. Such levels form the so-called characteristic curves $A_n(Q)$, which eventually approach the Landau levels as $Q\rightarrow\infty$.

\subsection{Applicability of the perturbation theory in $B$}\label{sectionF}
\begin{figure}
	\centering
	\includegraphics[width=0.8\linewidth]{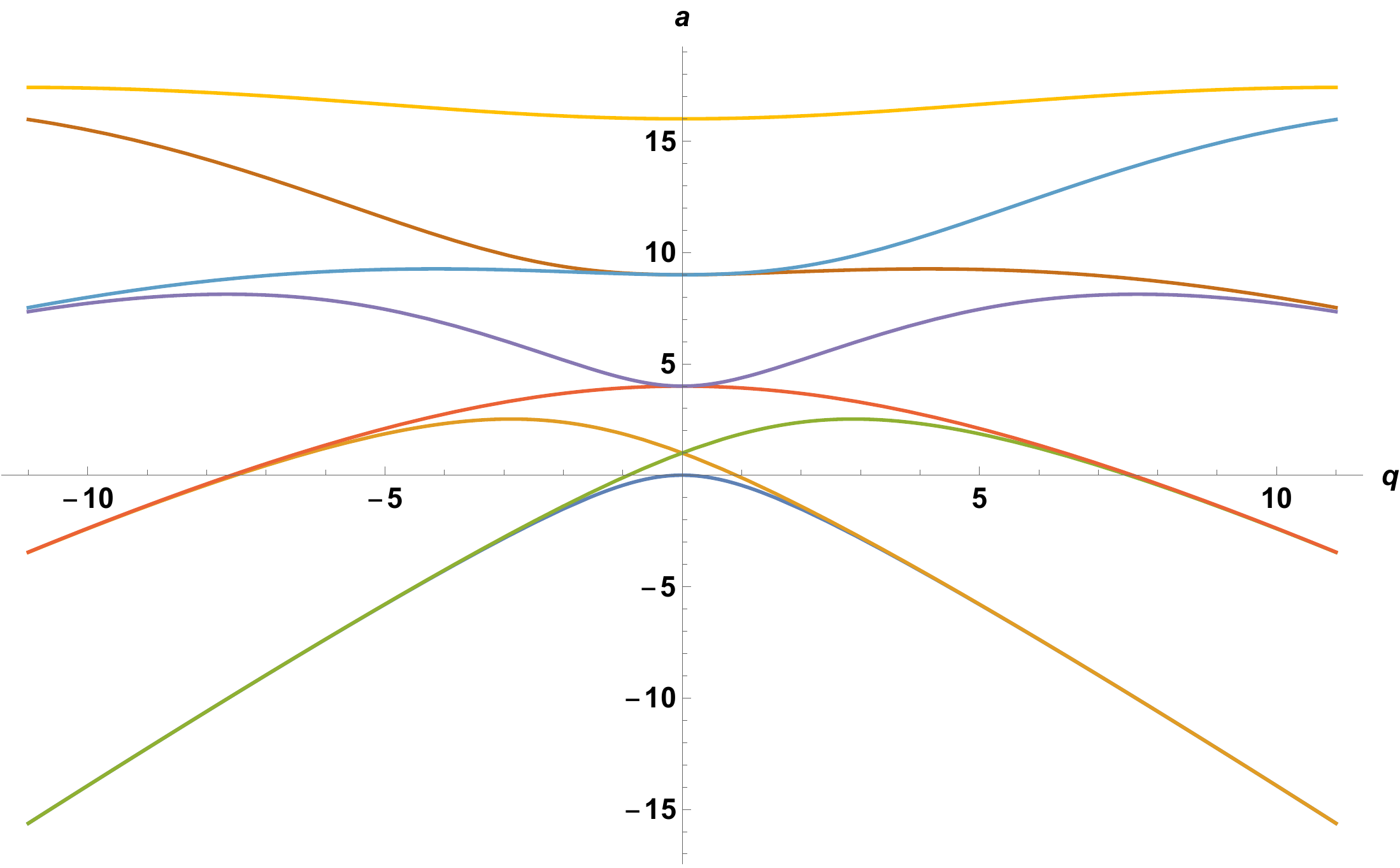}
	\caption{Finite-volume spectrum of the Mathieu equation.}
	\label{Mathieu_FV}
\end{figure}

The applicability of the main formula Eq. (\ref{finalresult}) is limited, since its derivation relies on a perturbative expansion of the energy shift in the external
field strength $B$. In Ref.~\cite{Agadjanov:2016cjc}, we have made a crude estimate for the upper bound on the magnitude of $B$ by considering a single period as a potential well. The condition that no bound states are formed in this potential well has led to the relation $eB<2\omega^2$. We are now in a position to provide a more stringent estimate, which is based on the properties of the solutions of the Mathieu equation.
To this end, we consider another limiting case, when $B\rightarrow 0$ while the frequency $\omega$ is fixed. Since $q=O(B)$ and $p=O(B^2)$, from 
Eq.~(\ref{WH}) we again arrive at the Mathieu equation:
\begin{equation}\label{M_eq}
\left[\frac{d^2}{dz^2}+a-2q\cos(2z) \right]f(z)=0\,.
\end{equation}
The structure of the finite-volume spectrum is shown in Fig. \ref{Mathieu_FV}. 
The expansion of  the ground state level $a_0$ in small $q$ reads \cite{AS}
\begin{eqnarray}\label{a0}
a_0=-\frac{q^2}{2}+O(q^4)\,,
\end{eqnarray}
A similar formula can be written by using exact power series for $a_0$ in small $p$ and $q$, see Ref. \cite{Urwin}. In particular, the first terms in these expansions coincide.

\begin{figure}
	\centering
	\includegraphics[width=0.8\linewidth]{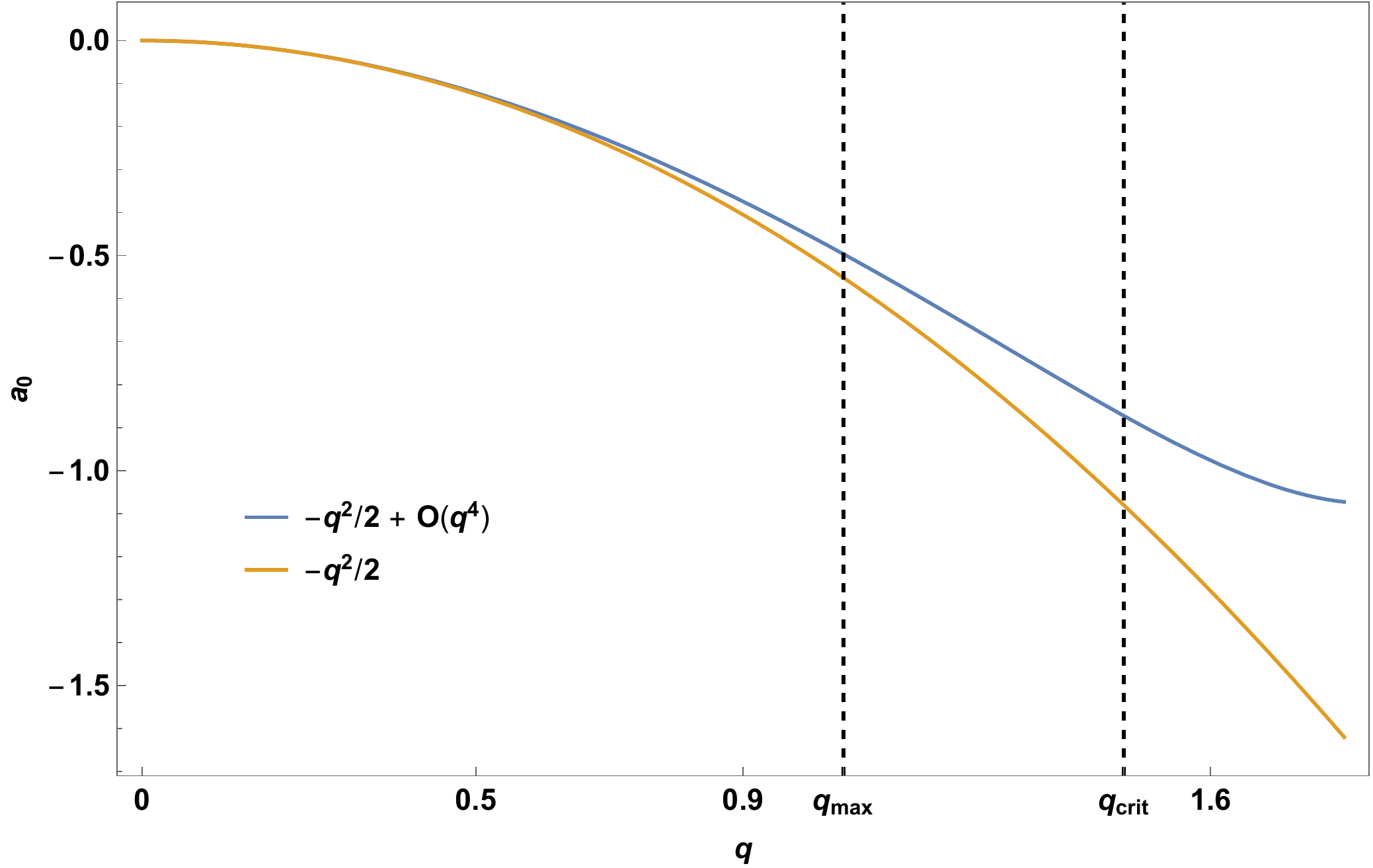}
	\caption{The ground state level $a_0(q)$ as a function of $q$. The critical value $q_{crit}\approx 1.47$ denotes the radius of convergence of the expansion in Eq. (\ref{a0}). At $q_{max}\approx 1.05$, the higher order corrections amount to a $10\%$ of the leading piece.} 
	\label{Mathieu_corr}
\end{figure}

The perturbative expansion of $a_0$ in
small $q$, Eq.~(\ref{a0}), has a certain finite radius of convergence $q=q_{crit}$. This critical value provides an upper bound on the field strength $B$. One obtains for the proton and neutron, respectively:
\begin{eqnarray}\label{B_up}
eB<\frac{q_{crit}\omega^2}{2(1+\kappa_p)}\,,\quad eB<\frac{q_{crit}\omega^2}{2|\kappa_n|}\,,
\end{eqnarray}
with $\kappa_p=1.79$ and $\kappa_n=-1.91$.
In case of the neutron, the numerical value of the radius of convergence reads $q_{crit}\approx 1.47$ (see also 
Ref.~\cite{Meixner}). For an estimate we use the same value for the proton. This is justified since in the limit $B\rightarrow 0$, Eq.~(\ref{WH}) can be well approximated by the Mathieu equation. Accordingly, one obtains $eB<0.26\omega^2$ (proton) and $eB<0.38\omega^2$ (neutron).

The upper bound on the magnetic field
strength in Eq.~(\ref{B_up}) can be improved by noting that our perturbative result, Eq. (\ref{finalresult}), might not be applicable at $q=q_{crit}$. To estimate higher order corrections, we can again resort to Eq.~(\ref{a0}). In Fig.~\ref{Mathieu_corr} we plot
the function $a_0(q)$  as well as the first term in Eq.~(\ref{a0}). As is seen,  the higher order terms become large at $q=q_{crit}$ and hence can not be neglected anymore. Accordingly, it is plausible to choose a certain value $q=q_{max}$, for which they, e.g., amount to a $10\%$ of the leading piece. This gives $q_{max}\approx 1.05$ (see also Fig.~\ref{Mathieu_corr}). Using 
Eq.~(\ref{B_up}) we get an improved bound on the magnitude $B$: $eB<0.19\omega^2$ and $eB<0.27\omega^2$ for the proton and the neutron, respectively.

It interesting to note that  Eq.~(\ref{a0}) allows us to verify the main result, Eq.~(\ref{finalresult}), in the approximation where the proton is treated as a point-like particle but with a non-zero anomalous magnetic moment. This is equivalent to setting $F_1=1$, $F_2=\kappa$ and $\beta_M=0$ in Eqs. (\ref{S1el},\ref{LET}). The subtraction function takes the value
\begin{eqnarray}\label{S1_point}
S_1(-\omega^2)=-\frac{1}{\omega^2}\left[(1+\kappa)^2-1 \right]\,. 
\end{eqnarray}
It is seen from Eq. (\ref{a0})  that there is no spin-dependent contribution to $a_0$. Using Eq.~(\ref{aqq'}), we get directly the spin-averaged energy shift:
\begin{eqnarray}
\delta E=-\frac{e^2B^2}{4m\omega^2}\left[(1+\kappa)^2-1 \right]\,,
\end{eqnarray}
where we have used the relation $E^2-m^2\approx 2m\delta E$. This is precisely the expression which is obtained from the main formula in Eq. (\ref{finalresult}), when we substitute the  subtraction function given in Eq.~(\ref{S1_point}).

The above discussion is equally applicable for the neutron, for which $g=0$.
In particular, the differential equation for $F(\m x)$, 
Eq.~(\ref{Deq_p}), simplifies:
\begin{eqnarray}\label{Deq_n}
[{\m p}^2-\kappa\mb\sigma\cdot \m B]F(\m x)=(E^2-m^2)F(\m x)\,.
\end{eqnarray}
It can be brought into the form of the Mathieu equation:
\begin{equation}\label{M_eqn}
\left[\frac{d^2}{dz^2}+a'-2q'\cos(2z) \right]f(z)=0\,,
\end{equation}
where 
\begin{equation}\label{a'q'}
a'=\frac{4}{\omega^2}\left( E^2-m^2\right)\,\quad q'=\alpha \frac{2\kappa eB}{\omega^2}\,.
\end{equation}
As expected, no Landau levels emerge in the zero frequency limit $\omega\rightarrow 0$. Further, the main formula in
Eq.~(\ref{finalresult}) can be verified in a similar manner by setting  $F_1=0$, $F_2=\kappa$ and $\beta_M=0$ in 
Eqs.~(\ref{S1el},\ref{LET}). The spin-averaged energy shift reads
\begin{eqnarray}
\delta E=-\frac{e^2B^2}{4m\omega^2}\kappa^2 \,.
\end{eqnarray}

\section{Propagator in an external field}
\label{sec:rel}

The derivation based on the non-relativistic framework, which was given in the
previous section, is not well suited for the study of the (exponentially suppressed)
finite-volume effects. For example, the final result, displayed in Eq.~(\ref{finalresult}),
contains the infinite-volume Compton amplitude on the right-hand side, i.e., the
finite-volume effects are neglected there. In the non-relativistic framework, these
effects may emerge from different sources. First, the non-relativistic couplings contain
the finite-volume corrections which, generally, go as $\exp(-M_\pi L)$ for large $L$ (we remind the reader that the lightest hadron mass gives the hard scale of the non-relativistic approach). Second,
 the Lagrangian contains operators which break rotational
invariance but preserve octahedral symmetry. These operators are multiplied by the
couplings that vanish exponentially for large values of $L$. Finally, in a finite volume,
one may construct a new type of gauge-invariant operators (the Wilson line), which
are absent in the infinite volume and whose contribution is also multiplied by
exponentially suppressed couplings. Matching to the Chiral Perturbation Theory (ChPT) with the external field in a finite
volume uniquely determines all these couplings. For a detailed discussion of these
issues we refer the reader to Refs.~\cite{Detmold:2006vu,Hu:2007eb,Hu:2007ts,Tiburzi:2007ep,Tiburzi:2008pa,Tiburzi:2014zva}.

From the above discussion it is clear that, in order to evaluate the finite-volume effects,
it is better to work directly with ChPT in a finite volume, abandoning the
non-relativistic framework, which has proven very convenient for discussing the
zero-frequency limit. The exponentially suppressed finite-volume effects, which
are {\em not} taken into account in ChPT, go as $\exp(-\Lambda_HL)$ instead of
$\exp(-M_\pi L)$ (here, $\Lambda_H$ denotes a typical hadronic scale of order of
one $\mbox{GeV}$), and thus can be neglected.

One important remark is in order. A procedure, which is used in 
Refs.~\cite{Detmold:2006vu,Hu:2007eb,Hu:2007ts,Tiburzi:2007ep,Tiburzi:2008pa,Tiburzi:2014zva} 
for the extraction of the polarizabilities in a finite volume, boils down to the 
derivation of the finite-volume one-particle effective action and to the
identification of the different terms in this action. In this
way, one again encounters the problem with  operators containing Wilson lines
that makes e.g., the authors of Ref.~\cite{Hu:2007ts} to conclude that ``At finite volume,
there is no longer a discernible relation between polarizabilities and the Compton tensor.''
In our framework we shall choose a different path, directly relating the amplitude for
forward Compton scattering in a finite volume to the second-order energy shift
of the nucleon in the external magnetic field.
We are not asking ourselves,
what the subtraction function
is in a finite volume -- this question anyway does not have an unique answer and, making
an inconvenient choice, one can easily obscure the relation between the
infinite- and finite-volume quantities. Rather, we can uniquely identify the quantity that
is extracted from the nucleon energy shift in a finite volume and which reduces to the
subtraction function in the limit $L\to\infty$ (as we shall see below, this is a certain
component of the spin-averaged Compton tensor in a particular kinematics). This fully
suffices to {\em define} a finite-volume counterpart of the subtraction function $S_1(q^2)$
and to calculate finite-volume corrections in an unambiguous way.

In this section, using  the framework of the effective field theory in a finite volume,
we shall derive the expression for the nucleon energy shift in an external field
(a finite-volume analog of Eq.~(\ref{finalresult})). Note also that we shall never specify
the Lagrangian of this theory -- it is only used to catalyze the proof and produce the
diagrammatic expansion of all amplitudes in terms of hadronic propagators.

We start from the nucleon two-point
function in the external field in  Minkowski space and define:
\begin{equation}
\tilde D(x,y)=i\langle 0|T\Psi(x)\bar\Psi(y)|0\rangle_A\,,
\end{equation}
where $\Psi(x)$ denotes the four-component spinor field, describing the nucleon.
Note that the Dirac indices are not shown explicitly.
Since the external field does not depend on time, we have $\tilde D(x,y)=
\tilde D(x^0-y^0;\m x,\m y)$. Further, it is convenient to define the Fourier transform
in the fourth component
\begin{equation}
\tilde D(\m x,\m y;E)=\int_{-\infty}^{\infty} dz^0 e^{iEz^0}\,
\tilde D(z^0;\m x,\m y)\, ,
\end{equation}
as well as in vector components,
\begin{equation}
D(\m p,\m k;E)=\int_{-L/2}^{L/2}d^3\m xd^3\m y\,e^{-i\m p\m x+i\m k\m y}\tilde D(\m x,\m y;E)\,.
\end{equation}
One can invert this expression, giving
\begin{equation}
\tilde D(\m x,\m y;E)=\frac{1}{L^3}\sum_{\m p}\frac{1}{L^3}\sum_{\m k}\,e^{i\m p\m x-i\m k\m y}D(\m p,\m k;E)\,.
\end{equation}

The free propagator takes the form:
\begin{equation}
D^{(0)}(\m p;\m k;E)=D^{(0)}(\m p;E)L^3\delta_{\m p\m k}\,,\quad\quad
 D^{(0)}(\m p;E)=\frac{m+\gamma_0E-\boldsymbol{\gamma}\m p}{m^2-E^2+\m p^2-i\epsilon}\,,
\end{equation}
where $m$ is the physical nucleon mass.

\begin{figure}
	\centering
	\includegraphics[width=0.95\linewidth]{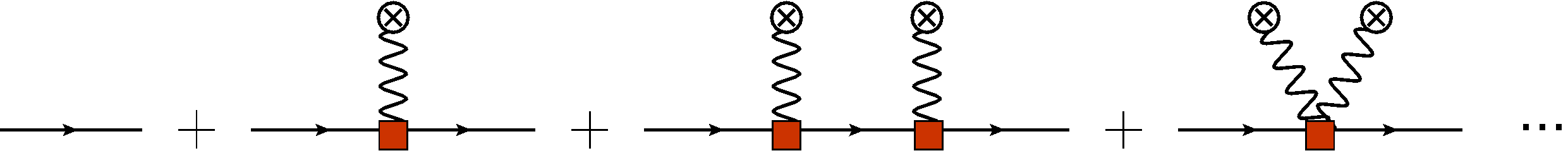}
	\caption{Diagrammatic representation of the propagator in an external 
                 electromagnetic field. The third and fourth diagrams correspond to 
                 the one-particle reducible and one-particle irreducible contributions 
                 at $O(B^2)$, respectively.
        }
	\label{propagator_ext}
\end{figure}

The diagrammatic representation of the propagator in the external electromagnetic field is 
schematically shown in Fig. \ref{propagator_ext}.  We have
\begin{eqnarray}\label{D_ext}\nonumber
D(\m p,\m k;E)&=& D^{(0)}(\m p;E)L^3\delta_{\m p\m k}
+D^{(0)}(\m p;E)\Sigma(\m p, \m k;E)D^{(0)}(\m k;E)\\[2mm]
&+&\frac{1}{L^3}\sum_{\m l}\,D^{(0)}(\m p;E)\Sigma(\m p, \m l;E)D^{(0)}(\m l;E)
\Sigma(\m l, \m k;E)D^{(0)}(\m k;E)+\cdots\,,
\end{eqnarray}
where the self-energy part $\Sigma(\m p, \m k;E)$, which is a matrix in the space of Dirac indices, 
can be expanded in powers of the magnitude of the external field:
\begin{eqnarray}
\Sigma(\m p, \m k;E)=\Sigma_0(\m p, \m k;E)+(eB)\Sigma_1(\m p, \m k;E)+(eB)^2\Sigma_2(\m p, \m k;E)+\cdots\,.
\end{eqnarray}
Here,  $\Sigma_0(\m p, \m k;E)=L^3\delta_{{\bf p}{\bf k}}\Sigma_0(\m p;E)$
is the sum of all one-particle irreducible diagrams in the absence of
the external field, and the functions $\Sigma_1,\Sigma_2$ will be determined below.
Note also that we use the MOM renormalization scheme, i.e., 
$\Sigma_0(\m p;\sqrt{m^2+{\bf p}^2})=0$.
The sum in Eq.~(\ref{D_ext}) can be written in a compact form:
\begin{eqnarray}\label{D_prop}
D(\m p,\m k;E)=D^{(0)}(\m p;E)L^3\delta_{\m p\m k}+D^{(0)}(\m p;E)T(\m p,\m k;E)D^{(0)}(\m k;E)\,,
\end{eqnarray}
where the amplitude $T(\m p,\m k;E)$ satisfies the relation
\begin{eqnarray}
T(\m p,\m k;E)=\Sigma(\m p, \m k;E)+\frac{1}{L^3}\sum_{\m l}\,
\Sigma(\m p, \m l;E)D^{(0)}(\m l;E)T(\m l,\m k;E)\,,
\end{eqnarray}
which is similar to the Lippmann-Schwinger equation.

In order to find the energy shift of the nucleon ground state, we have to determine
the pole position in the propagator $D(\m 0,\m 0;E)$. For this purpose, we single out the
term with ${\bf l}={\bf 0}$ in the sum (corresponding to the unperturbed ground state)
and rewrite 
the amplitude $T(\m 0,\m 0;E)$ as follows:
\begin{eqnarray}\label{T_alt}
T(\m 0,\m 0;E)=T'(\m 0,\m 0;E)+\frac{1}{L^3}T'(\m 0,\m 0;E)D^{(0)}(\m 0;E)T(\m 0,\m 0;E)\, ,
\end{eqnarray}
where the quantity $T'(\m p,\m k;E)$ satisfies the equation
\begin{eqnarray}
T'(\m p,\m k;E)=\Sigma(\m p, \m k;E)+\frac{1}{L^3}\sum_{\m l\neq 0}\,\Sigma(\m p, \m l;E)
D^{(0)}(\m l;E)T'(\m l,\m k;E)\,.
\end{eqnarray}
Note that now the sum runs over all $\m l\neq 0$. We get
\begin{eqnarray}
T(\m 0,\m 0;E)=\left( I-\frac{1}{L^3}T'(\m 0,\m 0;E)D^{(0)}(\m 0;E)\right)^{-1} T'(\m 0,\m 0;E)\,.
\end{eqnarray}
Here, $I$ denotes the unit $4\times 4$ matrix.
Inserting this expression into Eq.~(\ref{D_prop}) for the propagator $D(\m 0,\m 0;E)$, one obtains
\begin{eqnarray}
D(\m 0,\m 0;E)=D^{(0)}(\m 0;E)L^3\left( I-\frac{1}{L^3}T'(\m 0,\m 0;E)D^{(0)}(\m 0;E)\right)^{-1}\,.
\end{eqnarray}
Obviously, $D(\m 0,\m 0;E)$ and $T(\m 0,\m 0;E)$ have the same pole structure.

In order to simplify the matrix equation (\ref{T_alt}), we can use 
the octahedral symmetry of the cubic lattice. For zero momenta, the symmetry
requires that:
\begin{eqnarray}
R_{\alpha\gamma}(g)T_{\gamma\delta}(\m 0,\m 0;E)R_{\delta\beta}(g^{-1})=T_{\alpha\beta}(\m 0,\m 0;E)\,,
\end{eqnarray}
where $g$ denotes an arbitrary element of the octahedral group and $R_{\alpha\beta}(g)$ is 
the matrix of the linear representation of the octahedral group
which is obtained by restricting the $(1/2,0)+(0,1/2)$ representation of the 
Lorentz group to its octahedral subgroup (here the Greek letters denote
Dirac indices). The requirement of invariance restricts $T_{\alpha\beta}(\m 0,\m 0;E)$ to  the form:
\begin{eqnarray}
T_{\alpha\beta}(\m 0,\m 0;E)=\delta_{\alpha\beta}T^{(1)}+(\gamma_0)_{\alpha\beta}T^{(2)}\,,
\end{eqnarray}
where $T^{(1)}$ and $T^{(2)}$ are scalar functions. Accordingly, we see that 
\begin{eqnarray}
\bar u(\m 0,s)T(\m 0,\m 0;E) u(\m 0,s')&=& 2m\delta_{ss'}(T^{(1)}+T^{(2)})
\equiv2m\delta_{ss'}\tilde{T}\,,\\[2mm]
\bar v(\m 0,s)T(\m 0,\m 0;E) v(\m 0,s')&=&-2m\delta_{ss'}(T^{(1)}-T^{(2)})\,,\\[2mm]
\bar v(\m 0,s)T(\m 0,\m 0;E) u(\m 0,s')&=&\bar u(\m 0,s)T(\m 0,\m 0;E) v(\m 0,s')=0\,.
\end{eqnarray}
Similar relations can be established for the amplitude $T'(\m 0,\m 0;E)$. 
For example, $\tilde{T}'$ is defined through the equation
\begin{eqnarray}
\bar u(\m 0,s)T'(\m 0,\m 0;E) u(\m 0,s')=2m\delta_{ss'}\tilde{T}'\,.
\end{eqnarray}
Further, it is convenient to write the free propagator $D_{\alpha\beta}^{(0)}(\m p;E)$
in the form
\begin{eqnarray}
  D_{\alpha\beta}^{(0)}(\m 0;E)=
  \frac{1}{2m}\sum_{s}\frac{u_{\alpha}(\m 0,s)\bar u_{\beta}(\m 0,s)}{m-E}
  -\frac{1}{2m}\sum_{s}\frac{v_{\alpha}(\m 0,s)\bar v_{\beta}(\m 0,s)}{m+E}\,.
\end{eqnarray}
Multiplying the Eq.~(\ref{T_alt}) by $\bar u( \m 0,s)$ from the left and by $u(\m 0,s)$ from
the right, we get:
\begin{eqnarray}
\tilde{T}=\tilde{T}'+\frac{1}{L^3}\,\frac{1}{m-E}\tilde{T}'\tilde{T}\,.
\end{eqnarray}
This equation is not a matrix equation anymore. The pole position is given by
\begin{eqnarray}\label{eq:master}
m-E-\frac{1}{L^3}\tilde{T}'=0\,,
\end{eqnarray}
As expected the free pole at $m-E=0$ has disappeared. This result 
is similar to the ``master equation'' in case of hadronic 
atoms~\cite{Gasser:2007zt}.

Next, let us proceed with the calculation  of the amplitude $T'(\m 0,\m 0;E)$. In perturbation
theory, up-to-and-including $O(B^2)$ this quantity reads
\begin{eqnarray}\label{T'}
  T'(\m 0,\m 0;E)&=&\Sigma_0(\m 0, \m 0;E)+(eB)\Sigma_1(\m 0, \m 0;E)+(eB)^2\Sigma_2(\m 0, \m 0;E)\nonumber\\[2mm]
                 &+&(eB)^2\frac{1}{L^3}\sum_{\m l\neq 0}\,\Sigma_1(\m 0, \m l;E)D^{(0)}(\m l;E)
                     \Sigma_1(\m l, \m 0;E)+\cdots
                     \nonumber\\[2mm]
                 &=&T_0'(\m 0, \m 0;E)+(eB)T_1'(\m 0, \m 0;E)+(eB)^2T_2'(\m 0, \m 0;E)+\cdots\, .
\end{eqnarray}
The quantity $\Sigma_1(\m p, \m k;E)$ can be expressed through the three-point vertex function. 
Indeed, consider the linear coupling to the external field which is described by the Lagrangian
\begin{eqnarray}\label{eq:L_linear}
{\cal L}_1=-O_1(x)A^1(x)\, .
\end{eqnarray}
Here, we have used Eq.~(\ref{A-potential}) which states that the vector
potential has only one nonzero component $A^1$. The current operator $O_1$ contains both the 
nucleon and pion fields and obeys usual restrictions (hermiticity, certain transformation
properties with respect to the Lorentz group, etc.) but, otherwise, its form can be
arbitrary. Let us denote as $\tilde D_1(x, y)$ the respective contribution to
the two-point function. At order $O(B)$, we then have:
\begin{eqnarray}\label{D_1}
\tilde D_1(x,y)=\int_{-L/2}^{L/2}d^3 \m u\int_{-\infty}^{\infty}du_0\,A^1(\m u)\langle 0|
T\Psi(x)\bar\Psi(y)O_1(u)|0\rangle\,.
\end{eqnarray}
Further, using translational invariance, one can write
\begin{eqnarray}\label{mat_el}
\langle 0|T\Psi(x)\bar\Psi(y)O_1(u)|0\rangle
&=&\int_{-\infty}^{\infty}\frac{dp_0}{2\pi}\int_{-\infty}^{\infty}\frac{dk_0}{2\pi}
  \frac{1}{L^3}\sum_{\m p}\frac{1}{L^3}\sum_{\m k}\,e^{-i p (x-u)+ik (y-u)}
\nonumber\\[2mm]  
&\times&D^{(0)}( \m p,p_0 )\Gamma(p,k)D^{(0)}( \m k;k_0 )\,.
\end{eqnarray}
This is a definition of the vertex function $\Gamma(p;k)$. Substituting this expression into Eq.~(\ref{D_1}) and integrating over $u_0$ and $k_0$, we get
\begin{eqnarray}
\tilde D_1(x,y)&=&\int_{-L/2}^{L/2}d^3 \m u\,A^1(\m u)\int_{-\infty}^{\infty}\frac{dp_0}{2\pi}\frac{1}{L^3}\sum_{\m p}\frac{1}{L^3}\sum_{\m k}\,e^{-i p_0 (x_0-y_0)}e^{i \m p (\m x-\m u)-i\m k (\m y-\m u)}
\nonumber\\[2mm]
&\times& D^{(0)}( \m p;p_0 )\Gamma(\m p,\m k;E)D^{(0)}( \m k,p_0 )\,,
\end{eqnarray}
where $\Gamma(\m p,\m k;E)$ is obtained from $\Gamma(p,k)$ by substituting $p_0=k_0=E$ (we remind the reader that the field $A^1$ is static).
Accordingly, the Fourier transform of $\tilde D_1(x,y)$ takes the simple form:
\begin{eqnarray}
D_1(\m p,\m k;E)=\int_{-L/2}^{L/2}d^3 \m u\,A^1(\m u)e^{-i(\m p-\m k)\m u}D^{(0)}( \m p;E )\Gamma(\m p,\m k;E)D^{(0)}( \m k;E )\,.
\end{eqnarray}
Comparing this result with the expansion of the propagator in Eq.~(\ref{D_ext}) at
$O(B)$, it is seen that
\begin{eqnarray}
(eB)\Sigma_1(\m p,\m k;E)=\Gamma(\m p,\m k;E)
  \int_{-L/2}^{L/2}d^3 \m u\,A^1(\m u)e^{-i(\m p-\m k)\m u}\doteq
  \Gamma(\m p,\m k;E) \tilde A^1({\bf k}-{\bf p})\,.
\end{eqnarray}

Next, we evaluate the quantity $\Sigma_2(\m p, \m k;E)$ which consists of all one-particle
irreducible diagrams with amputated nucleon legs, with two external fields attached.
Let $\Upsilon(p,k,l)$ denote the sum of all such diagrams in momentum space. Here,
$p$ and $k$ denote the momenta of the outgoing and ingoing nucleon, respectively,
and the momenta of two external ``photons'' are equal to $l+(k-p)/2$ and $l-(k-p)/2$,
respectively. Further, denoting
\eq
\Upsilon({\bf p},{\bf k},{\bf l};E)=\Upsilon(p,k,l)\biggr|_{p_0=k_0=E,l_0=0}\, ,
\en
it is easy to check that $\Sigma_2$ is given by
\begin{eqnarray}
  (eB)^2\Sigma_2(\m p, \m k;E)&=&\int_{-L/2}^{L/2}d^3 \m u\int_{-L/2}^{L/2}d^3 \m v\,A^1(\m u)
                                  A^1(\m v)
\nonumber\\[2mm]
  &\times&\frac{1}{L^3}\sum_{\m l}\,
  e^{i \m u (\m l+(\m k-\m p)/2)-i\m v(\m l-(\m k-\m p)/2)}\Upsilon(\m p,\m k,\m l;E)\,.
\end{eqnarray}

 \begin{figure}
 	\centering
 	\includegraphics[width=0.45\linewidth]{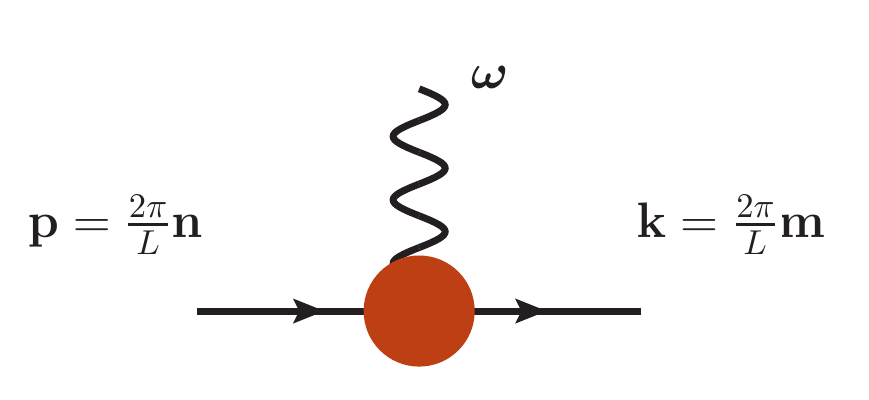}
 	\caption{Three-momentum conservation in the $\gamma^*NN$ vertex gives the Kronecker 
                 delta $\delta_{\m p-\m k ,\pm \mb \omega}$.  }
 	\label{fig:3pvertex}
 \end{figure}

We now have all ingredients for the evaluation of the energy shift.
Here, following the discussion in the previous section, we consider the scenario a) for 
the external field, when the frequency $\omega$ is quantized and the three-momentum 
conservation holds. In this case, the term linear in $B$ gives:
\begin{eqnarray}
  (eB)T_1'(\m p, \m k;E)=(eB)\Gamma(\m p,\m k; E)\frac{L^3}{2i\omega}[
  \delta_{-\m p+\m k,-\boldsymbol \omega}-\delta_{-\m p+\m k,\boldsymbol\omega}]\,.
\end{eqnarray}
Obviously, $\Sigma_1(\m 0, \m 0;E)=0$ for $\boldsymbol\omega\neq {\bf 0}$, see also 
Fig.~\ref{fig:3pvertex}.

The second-order term at threshold takes the form
\begin{eqnarray}\label{T'I}
(eB)^2T_2'(\m 0,\m 0;m)&=&
 \frac{(eB)^2L^3}{4\omega^2}[\Upsilon(\m 0,\m 0, \boldsymbol{\omega};m)
 +\Upsilon(\m 0,\m 0,-\boldsymbol{\omega};m)]
                     \nonumber\\[2mm]
                 &+&\frac{(eB)^2L^3}{4\omega^2}\,\biggl\{
                     \Gamma(\m 0,\boldsymbol\omega; m)
                     D_0(\boldsymbol\omega,m)\Gamma(\boldsymbol\omega,\m 0; m)
\nonumber\\[2mm]
                 &+&\Gamma(\m 0,-\boldsymbol\omega; m)
                     D_0(-\boldsymbol\omega,m)\Gamma(-\boldsymbol\omega,\m 0; m)
                     \biggr\}\,.
\end{eqnarray}
On the other hand, it is straightforward to verify that the expression
on the right-hand side of this equation is proportional to the ``11'' component
of the forward Compton scattering tensor. Note also that, since $\boldsymbol\omega\neq{\bf 0}$, 
the sum over the intermediate nucleon states does not contain the term with
${\bf l}={\bf 0}$ and hence, there is no difference between $T_2$ and $T_2'$ at threshold.
Thus
\begin{eqnarray}
  \frac{Z(eB)^2}{2}\,\sum_s\bar u({\bf 0},s)T_2'(\m 0,\m 0;m)u({\bf 0},s)
  =\frac{(eB)^2L^3}{2\omega^2}\,T^{11}(p,q)\, ,
\end{eqnarray}
where $p^\mu=(m,{\bf 0})$, $q^\mu=(0,{\boldsymbol\omega})$ and the nucleon wave
function renormalization constant (in the absence of the external field) is given by
\eq
Z^{-1}&=&1+\frac{d}{dE}\, \tilde\Sigma_0(\m 0;E)\biggr|_{E=m}\, ,\quad\quad
2m\delta_{ss'}\tilde\Sigma_0(\m 0;E)=\bar u({\bf 0},s)\Sigma_0(\m 0;E)u({\bf 0},s')\,.
\en
It is now straightforward to determine the spin-averaged energy shift at order $B^2$
from Eq.~(\ref{eq:master}). Taking into account the fact that the linear term in $B$
vanishes, and expanding the quantity $\tilde T'$ in Eq.~(\ref{eq:master}) in Taylor
series in $E-m$, it immediately follows that the spin-averaged energy shift is given by
\begin{eqnarray}
\delta E
  &=&-\frac{Z(eB)^2}{2mL^3}\frac{1}{2}\sum_{s}\,\bar u(\m 0,s)T_2'(\m 0,\m 0;E)u(\m 0,s) +O(B^3)\, ,
 \end{eqnarray}
from which we finally obtain
\begin{equation}\label{eq:finalresult-L}
\delta E=-\frac{1}{4m}\,\left( \frac{eB}{\omega}\right)^2 T^{11}(p,q)+O(B^3)\,.
\end{equation}
This equation is the finite-volume version of Eq.~(\ref{finalresult}) and contains the
``11'' component of the Compton tensor, evaluated in a finite volume. Up to the
corrections, proportional to $\exp(-M_\pi L)$, this quantity is given by the subtraction
function $S_1(q^2)$ via $T^{11}(p,q)=-\omega^2S_1(q^2)$, see Eqs.~(\ref{Taveraged})
and (\ref{eq:K}).
Hence, the equation~(\ref{eq:finalresult-L}) provides a framework for the
systematic calculation of such corrections.

\section{Discussion of the parameter range in numerical calculations}
  In Ref.~\cite{Agadjanov:2016cjc} we  presented a brief discussion of the lattice
  parameters, which could be used in the numerical extraction of the subtraction function $S_1(q^2)$.
  With the use of the new, more stringent constraints on the value of the magnetic field
  we are now able to refine this analysis. For a moment, we neglect the finite-volume
  corrections altogether. As in Ref.~\cite{Agadjanov:2016cjc}, the elastic
  and inelastic parts of the amplitude are parameterized as:
  \eq
  S_1^{el}(q^2)&=&-\frac{4m^2}{q^2(4m^2-q^2)}\,\biggl\{G_E^2(q^2)-G_M^2(q^2)\biggr
  \}\, ,
  \nonumber\\[2mm]
  S_1^{inel}(q^2)&=&S_1^{inel}(0)G_d(q^2)\, ,\quad\quad
  S_1^{inel}(0)=-\frac{\kappa^2}{4m^2}-\frac{m}{\alpha}\,\beta_M\, ,
  \en
  where $\kappa$ and $\beta_M$ denote the anomalous magnetic moment and the
  magnetic polarizability of the nucleon (we use the same numerical values in the estimates
  as given in Ref.~\cite{Agadjanov:2016cjc}).
  Further, $G_d(q^2)=(1-q^2/0.71\mbox{\,GeV}^2)^{-2}$ is the dipole form factor. It should be noted that the asymptotic behavior at large values of $q^2$ is consistent with the
  result of the operator product expansion in QCD.

  The electric and magnetic form factors of the proton and the neutron are given by:
  \eq
  G_E^p(q^2)&=&G_d(q^2)\, ,\quad\quad G_M^p(q^2)=(1+\kappa^p)G_d(q^2)\,,
  \nonumber\\[2mm]
  G_E^n(q^2)&=&\frac{-q^2}{4m^2}\,\kappa^nG_d(q^2)\, ,\quad\quad
  G_M^n(q^2)=\kappa^nG_d(q^2)\,,
  \en
  with the same dipole form factor as above.
  
  One of the estimates, which goes through exactly in the same way as in
  Ref.~\cite{Agadjanov:2016cjc}, is related to our ability to separate the physically
  interesting inelastic part from total amplitude. As seen, the elastic part is singular
  at threshold and falls off very fast at higher $\omega^2$. Thus, the separation will be
  difficult for very small values of $\omega^2$. One may require, for instance, that at
  the minimum value of $\omega^2$, the inelastic contribution amounts
  up to a $10\%$ of the elastic contribution. In this manner, we get
  $\omega_{min}^2=0.086~\mbox{GeV}^2$ for the proton and 
  $\omega_{min}^2=0.045~\mbox{GeV}^2$ for the neutron (a slight difference to the numbers given in Ref.~\cite{Agadjanov:2016cjc} is caused by the fact that here we take into account the exact momentum dependence of all amplitudes). 
  If one requires instead that elastic and inelastic parts are equal,
  one gets $\omega_{min}^2=0.40~\mbox{GeV}^2$ for the proton and
  $\omega_{min}^2=0.26~\mbox{GeV}^2$ for the neutron. In any case, the lower cutoff
  on the available frequencies is rather comfortable and does not put significant
  restrictions on the parameters of the lattices which can be used in the calculations.
  
  The conditions, which involve the magnitude of the magnetic field, are more
  restrictive. On one side, the magnetic field should be strong enough, in
  order to measure the effect at all. On the other hand, it must be weak enough, so
  that the perturbation theory still applies. In section \ref{sectionF} we have made a more stringent estimate
  \eq
  eB<0.19\,\omega^2\quad\mbox{(proton),}\quad\quad
  eB<0.27\,\omega^2\quad\mbox{(neutron),}
  \en
  which is based on the properties of the
  solutions of Mathieu's equation. Denoting the inelastic shift by $\delta E^{inel}$,
  according to Ref.~\cite{Agadjanov:2016cjc}, we get
  \eq\label{ineq}
  eB&=&\biggl(\frac{4m\delta E^{inel}}{S_1^{inel}(0)}\biggr)^{1/2}G_d^{-1/2}(-\omega^2)<0.19\omega^2\quad\mbox{(for proton),}
  \nonumber\\[2mm]
  eB&=&\biggl(\frac{4m\delta E^{inel}}{S_1^{inel}(0)}\biggr)^{1/2}G_d^{-1/2}(-\omega^2)<0.27\omega^2\quad\mbox{(for neutron).}
  \en
  It is now clear that the window for the available values of $eB$ exists if and only if
  $\delta E^{inel}$ can be taken sufficiently small or, in other words, if the uncertainty in
  the determination of $\delta E^{inel}$ does not exceed certain value. Generously
  allowing this uncertainty to be $\delta E^{inel}=0.05m$, as done in Ref. \cite{Agadjanov:2016cjc}, is no more an option - the inequalities in Eq. (\ref{ineq}) can not be satisfied.
  A better accuracy in the determination of the energy shift $\delta E^{inel}=0.01m$
  would lead to the lower cutoff on the available frequencies
  $\omega_{min}^2=0.90~\mbox{GeV}^2$  for the proton and
  $\omega_{min}^2=0.41~\mbox{GeV}^2$ for the neutron. If one wants to increase the range of available frequencies and reach lower values of $\omega^2$,
  one has to improve on the accuracy further. The range of the magnitudes for the magnetic
  field can be determined from the above equations, if the accuracy is given.

\section{Conclusions and Outlook}

\begin{itemize}

\item[i)]
  We have presented three alternative derivations (the third one is contained in
  the appendix) of the formula for the energy shift of
  the nucleon, placed in a periodic external field. Namely, the non-relativistic effective
  Lagrangian was used for this purpose, as well as the relativistic framework. All these
  alternative settings are advantageous for discussing different issues arising in
  the treatment of the problem. The aim of the whole exercise is to extract the
  forward Compton scattering amplitude in a certain kinematics, the so-called subtraction
  function $S_1(q^2)$, from lattice simulations. In its turn,
  measuring this function would enable
  one to gain important information about the properties of QCD at low energy. 
  
\item[ii)]
  The result or Ref.~\cite{Agadjanov:2016cjc} has been refined and extended in various
  aspects. For example, in this paper we discuss in detail the zero-frequency limit
  (constant magnetic field)
  of the expression for the energy shift. This limiting case is studied in the 
  literature in detail.
  Here, it is shown that, to have a smooth transition
  to this limit, the external field on the lattice should be implemented in a specific way,
  corresponding to the scenario b). There exists no zero-frequency limit for scenario a).
\item[iii)]
  Another important issue is the limit of validity of the perturbative treatment of
  the external magnetic field (the convergence radius of the perturbative expansion in $B$).
  In Ref.~\cite{Agadjanov:2016cjc}, using heuristic arguments,
  we gave a rough estimate of the maximal field strength, for which the perturbative
  approach is still applicable. In the present paper we improved the argument and
  give a new, much more stringent estimate,
which is based on the properties of the
  solutions of Mathieu's equation. It should be pointed out that the non-relativistic EFT
  approach provides the most convenient framework for the discussion of the above two
  problems.

\item[iv)]
   We have generalized the result of
  Ref.~\cite{Agadjanov:2016cjc} and derived an equation which relates the energy shift to
  the ``11'' component of the Compton tensor in a finite volume. Using this formula,
  one may estimate the exponentially suppressed finite-volume corrections to
  the extracted value of the subtraction function. This can be done, e.g., by performing
  calculations at one loop in ChPT in a finite volume. The calculations are under way
  and the results will be reported elsewhere~\cite{Lozano}. The preliminary
  results show that the finite-volume corrections at one loop in ChPT are sizable, but
  can be kept under control at reasonably large lattice volumes $M_\pi L> 4$.
  Moreover, one might use
  a similar setting to estimate the effects of  partial (electro)quenching in lattice simulations.

\end{itemize}

\begin{acknowledgments}
	We thank  Z. Davoudi, M. Petschlies, M. Savage, G. Schierholz and B. Tiburzi
	for useful discussions. 
	We acknowledge the support from the DFG (CRC 110 ``Symmetries and the Emergence 
        of Structure in QCD'' 
	and  Bonn-Cologne Graduate School of Physics and
	Astronomy). This research is supported in part by Volkswagenstiftung
	under Contract No. 93562, by the Chinese Academy of Sciences (CAS) 
	President's
	International Fellowship Initiative (PIFI) (Grant No. 2018DM0034) and by Shota Rustaveli National Science Foundation
(SRNSF), grant no. DI-2016-26.

\end{acknowledgments}

\appendix*
\section{Alternative derivation of Eq. (\ref{eq:finalresult-L}).}
\label{app:A}
For the sake of completeness, we present yet another derivation of the main
formula for the energy shift given in Eq. (\ref{eq:finalresult-L}), which is based on the study of
the two-point function of the nucleon field in the external field at large time separation
and, hence, has a closer resemblance to the methods used on the lattice.
It is assumed that
the external field is implemented according to scenario a), i.e., the frequency is quantized. 
In this derivation,
we directly expand the nucleon two-point function in the external field
\begin{equation}\label{C_expansion}
C=C^{(0)}+C^{(1)}+C^{(2)}+O(A^3),
\end{equation}
where 
\begin{eqnarray}
C&=&\frac{1}{L^3}\int_{-L/2}^{L/2}d^3{\bf x}d^3{\bf y}\,\langle 0|T\Psi(x)\bar\Psi(y)|0\rangle_A\,,\\
C^{(0)}&=&\frac{1}{L^3}\int_{-L/2}^{L/2}d^3{\bf x}d^3{\bf y}\,\langle 0|T\Psi(x)\bar\Psi(y)|0\rangle\,,\\
C^{(1)}&=&\frac{i}{L^3}\int_{-L/2}^{L/2}d^3{\bf x}d^3{\bf y}d^4z\,A_\mu(\m z)\langle 0|T\Psi(x)\bar\Psi(y)j^\mu(z)|0\rangle\,,\\
C^{(2)}&=&\frac{i^2}{2L^3}\int_{-L/2}^{L/2}d^3{\bf x}d^3{\bf y} d^4zd^4v\,A_\mu(\m z)A_\nu(\m v)\langle 0|T\Psi(x)\bar\Psi(y)j^\mu(z)j^\nu(v)|0\rangle\,.
\end{eqnarray}
Here, the integration over $d^3{\bf x}d^3{\bf y}$ projects onto the states with zero initial and final three-momenta.

Note that, strictly speaking,
for a rigorous derivation one should perform the Wick rotation into
the Euclidean space and pick up the leading terms in the two-point function at large
(Euclidean) times. For simplicity, however, we stay in the Minkowski space and identify
the leading exponentials there -- in our case, the identification is easy and no
ambiguities arise.

The completeness condition, which we shall be using, takes the form
\begin{equation}
\frac{1}{L^3}\sum_{\m ks}\,\frac{|\m k, s\rangle\langle \m k, s|}{2\omega(\m k)}+\cdots=1\,,
\end{equation}
where ellipses stand for the excited states contributions; they will be neglected altogether by taking the limit $x_0-y_0\to\infty$ (the time extent of the lattice is assumed to be infinite). 

Let us start with the  matrix element $C^{(0)}$ that describes the propagation of the nucleon in the absence of the external field. Using the translation invariance, it takes the form
\eq
C^{(0)}&=&\frac{1}{L^6}\sum_{\m ks}\frac{1}{2w(\m k)} \int_{-L/2}^{L/2}d^3\m xd^3\m y\,\theta(x_0-y_0)e^{i\m k(\m x-\m y)}e^{-i w(\m k)( x_0- y_0)}\langle 0|\Psi(0)|\m k, s\rangle\langle \m k, s|\bar\Psi(0)|0\rangle
\nonumber\\[2mm]
&+&\cdots\,.
\en
Note that the second term in the $T$-product, containing $\theta(y_0-x_0)$, picks up the
antiparticle pole for large $x_0-y_0$ instead of the particle pole, and thus can  be neglected.

Integrating over all variables, one gets
\begin{equation}
  C^{(0)}=\sum_s\frac{e^{-im(x_0-y_0)}}{2m}\,
  \langle 0|\Psi(0)|\m 0, s\rangle\langle \m 0, s|\bar\Psi(0)|0\rangle+\cdots\, .
  \end{equation}
Taking into account that
\eq
\langle0|\Psi(0)|{\bf 0},s\rangle=Z^{1/2}u({\bf 0},s)\, ,
\en
where $Z$ denotes the wave function renormalization constant, we may write
\eq
\bar u({\bf 0},s)C^{(0)}u({\bf 0},s)=2mZe^{-im(x_0-y_0)}+\cdots\, .
\en

The two-point function in the presence of the external magnetic field can be written in a similar manner:
\begin{equation}\label{C_B}
  \bar u({\bf 0},s)Cu({\bf 0},s)=2E_s(B)Z_s(B)e^{-iE_s(B)(x_0-y_0)}+\cdots
 \end{equation}
  Here, $E_s(B)$ is the energy of the nucleon ground state. Note that, in general,
  $E_s(B)$ and $Z_s(B)$ depend on the orientation of the spin $s$. The functions $E_s(B)$ and $Z_s(B)$  can be expanded in $B$:
\begin{eqnarray}\label{Energy_B}
  E_s(B)&=&m+\xi_s(\omega) (eB)+\eta_s(\omega) (eB)^2+O(B^3)\,,
            \nonumber\\[2mm]
  Z_s(B)&=&Z+\alpha_s(\omega) (eB)+\beta_s(\omega) (eB)^2+O(B^3)\, .
\end{eqnarray}
The unknown quantities $\xi_s(\omega),\eta_s(\omega),\alpha_s(\omega),$ and $\beta_s(\omega) $ depend on the frequency $\omega$.

The correlator $C^{(1)}$ is evaluated in a similar manner to $C^{(0)}$. The only contribution remaining at $x_0-y_0\to\infty$ is given by
\begin{eqnarray}\nonumber
C^{(1)}&=&\frac{i}{L^9}\sum_{\m ks,\m ls'}\frac{1}{4w(\m k)w(\m l)}\int_{-L/2}^{L/2}d^3\m xd^3\m yd^3\m zdz_0\,A^1(\m z)\theta(x_0-z_0)\theta(z_0-y_0)\\[0.5cm]
&\times& e^{i\m k(\m x-\m z)}e^{-i w(\m k)( x_0- z_0)}e^{i\m l(\m z-\m y)}e^{-iw(\m l)( z_0- y_0)}
         \langle 0|\Psi(0)|\m k, s\rangle\langle \m k,s|j^1(0)|\m l, s'\rangle\langle \m l, s'|\bar\Psi(0)|0\rangle+\cdots\, .
         \nonumber\\
\end{eqnarray}
After the summation over the three-momentum the above expression
simplifies and we get:
\begin{eqnarray}
  \bar u({\bf 0},s)C^{(1)}u({\bf 0},s)=i\frac{Ze^{-im(x_0-y_0)}}{L^3}
  (x_0-y_0)\langle \m 0,s|j^1(0)|\m 0, s\rangle\int_{-L/2}^{L/2}d^3\m z\,A^1(\m z)+\cdots\,,
\end{eqnarray}
The integral over the periodic electromagnetic potential vanishes, and so 
\begin{equation}
C^{(1)}=0.
\end{equation}

The calculation of the second-order matrix element $C^{(2)}$ proceeds similarly.
First inserting the completeness relation and using the translational invariance, one gets
\begin{eqnarray}
\bar u({\bf 0},s)C^{(2)}u({\bf 0},s)&=&i^2\frac{Ze^{-im(x_0-y_0)}}{2L^3}\int_{-L/2}^{L/2}d^3\m zd^3\m vd\lambda_0dv_0\,A^1(\m z)A^1(\m v)
\theta(x_0-\lambda_0-v_0)\theta(v_0-y_0)
\nonumber\\[2mm]
&\times&\langle \m 0,s|Tj^1(\lambda)j^1(0)|\m 0, s\rangle+\cdots\,,
\end{eqnarray}
where $\lambda_0=z_0-v_0$  is a new integration variable,
and we have introduced the new four-vector
$\lambda=(\lambda_0,\m z-\m v)$. It is then straightforward to verify the identity
\begin{eqnarray}
\langle \m 0,s|Tj^1(\lambda)j^1(0)|\m 0, s\rangle=-\frac{2i}{L^3}\sum_{\m q}\int_{-\infty}^{\infty}\,\frac{dq_0}{2\pi}e^{-iq\lambda}T^{11}({\bf 0},s;{\bf 0},s;q)\,, \quad \lambda=z-v\,.
\end{eqnarray}
Here, $T^{11}$ is the ``11'' component of the Compton tensor (before spin averaging)
in a finite volume. We further get
\begin{eqnarray}\nonumber
  \bar u({\bf 0},s)C^{(2)}u({\bf 0},s)&=&i\frac{Ze^{-im(x_0-y_0)}}{L^6}\,
                                          \sum_{\m q}\int_{-\infty}^{\infty}\,\frac{dq_0}{2\pi}\int_{-L/2}^{L/2}d^3\m zd^3\m vd\lambda_0dv_0\\[0.2cm]
&\times& A^1(\m z)A^1(\m v)
\theta(x_0-\lambda_0-v_0)\theta(v_0-y_0)e^{-iq\lambda}
T^{11}({\bf 0},s;{\bf 0},s;q)+\cdots\,.\qquad
\end{eqnarray}
Next, the integration over $v_0$ gives
\begin{equation}
\int_{-\infty}^\infty  dv_0\,\theta(x_0-\lambda_0-v_0)\theta(v_0-y_0)=(x_0-y_0-\lambda_0)\theta(x_0-y_0-\lambda_0)\,.
\end{equation}
Accordingly,
\begin{eqnarray}\label{C2_sum}
  \bar u({\bf 0},s)C^{(2)}u({\bf 0},s)=i\frac{Ze^{-im(x_0-y_0)}}{L^6}\,
  \sum_{\m q}\,\tilde{A}^1(\m q)\tilde{A}^1(-\m q)I({\bf 0},s;{\bf 0},s;\m q)+\cdots\,,
\end{eqnarray}
where $\tilde{A}^1(\m q)$ is defined in Eq. (\ref{Ap}). The quantity
$I({\bf 0},s;{\bf 0},s;\m q)$ reads
\begin{equation}
I({\bf 0},s;{\bf 0},s;\m q)=\int_{-\infty}^{\infty}\,\frac{dq_0}{2\pi}\int_{-\infty}^{x_0-y_0}\,d\lambda_0\,(x_0-y_0-\lambda_0)e^{-iq_0\lambda_0}T^{11}({\bf 0},s;{\bf 0},s;q)\,.
\end{equation}
The shift of the variable $\lambda_0\rightarrow x_0-y_0-\lambda_0$ and partial integration over $q_0$ gives
\begin{equation}
I({\bf 0},s;{\bf 0},s;\m q)=i\int_{-\infty}^{\infty}\,\frac{dq_0}{2\pi}\int_{0}^{\infty}\,d\lambda_0\,e^{iq_0\lambda_0}\frac{\p}{\p q_0}\left[e^{-iq_0(x_0-y_0)} T^{11}({\bf 0},s;{\bf 0},s;q)\right]\,.
\end{equation}
Integrating over $\lambda_0$, one obtains
\begin{equation}
  I({\bf 0},s;{\bf 0},s;\m q)=-\int_{-\infty}^{\infty}\,\frac{dq_0}{2\pi}\frac{e^{-iq_0(x_0-y_0)}}{q_0+i\epsilon}\left[-i(x_0-y_0) T^{11}({\bf 0},s;{\bf 0},s;q)
    +\frac{\p}{\p q_0}T^{11}({\bf 0},s;{\bf 0},s;q)\right] \,,
\end{equation}
where the $i\epsilon$ prescription ensures the convergence of the integral. Further,
contour integration leads to the following expression:
\begin{equation}
  I({\bf 0},s;{\bf 0},s;\m q)=(x_0-y_0) T^{11}({\bf 0},s;{\bf 0},s;\bar q)
  +i\frac{\p}{\p q_0}T^{11}({\bf 0},s;{\bf 0},s; q)\Big|_{q=\bar q} \,,\quad \bar q=(0,\m q)\,.
\end{equation}
Finally, inserting this result into Eq. (\ref{C2_sum}) and summing over $\m q$, the correlator $C^{(2)}$ takes the value
\begin{eqnarray}
  \bar u({\bf 0},s)C^{(2)}u({\bf 0},s)&=& i\frac{Ze^{-im(x_0-y_0)}}{2}
  \left( \frac{eB}{\omega}\right)^2\bigg[(x_0-y_0) T^{11}({\bf 0},s;{\bf 0},s;\hat q)
\nonumber\\[2mm]
  &+& \frac{i}{2}\frac{\p}{\p q_0}T^{11}({\bf 0},s;{\bf 0},s; q)\Big|_{q=\hat q}
  +\frac{i}{2}\frac{\p}{\p q_0}T^{11}({\bf 0},s;{\bf 0},s;q)\Big|_{q=-\hat q}\bigg] +\cdots\, ,
\end{eqnarray}
where the symmetry property of the Compton tensor, $T^{11}({\bf 0},s;{\bf 0},s;\hat q)=T^{11}({\bf 0},s;{\bf 0},s;-\hat q)$, $\,\hat q=(0,\mb \omega)$, was used.

Next, combining Eqs. (\ref{Energy_B}) and (\ref{C_B}), one gets the Taylor expansion of the two-point function in the magnetic field strength:
\begin{eqnarray}
  \bar u({\bf 0},s)Cu({\bf 0},s)&=&2mZe^{-im(x_0-y_0)}
  \biggl\{ 1+\frac{\alpha_s(\omega)(eB)+\beta_s(\omega)(eB)^2}{Z}+\frac{\xi_s(\omega)(eB)+\eta_s(\omega)(eB)^2}{m}
 \nonumber\\[2mm]
  &-&i(x_0-y_0)\left.\left[
      \frac{\xi_s(\omega) (eB)+\eta_s(\omega) (eB)^2}{m}\right]+\frac{\xi_s(\omega)\alpha_s(\omega)(eB)^2}{mZ}\right.
 \nonumber\\[2mm]
&-&i(x_0-y_0)\frac{\xi_s(\omega)(eB)}{m}\,\left[\frac{\xi_s(\omega)(eB)}{m}+\frac{\alpha_s(\omega)(eB)}{Z}\right]
    +O(B^3,(x_0-y_0)^2)\biggr\} \, .
    \nonumber\\
\end{eqnarray}
The unknown coefficients $\xi_s(\omega)$ and $\eta_s(\omega)$ are determined from a comparison of the right-hand side in the above expansion with the known expression
of $C$ up to and including $O(B^2)$. This comparison gives the familiar result:
\begin{eqnarray}
\xi_s(\omega)=0,\qquad \eta_s(\omega)=-\frac{1}{4m\omega^2}T^{11}({\bf 0},s;{\bf 0},s;\hat q)\,.
\end{eqnarray} 
The quantities $\alpha_s(\omega)$ and $\beta_s(\omega)$ can be found in a similar fashion. In particular, $\alpha_s(\omega)=0$, while $\beta_s(\omega)$ is given as a certain linear combination of the tensor component $T^{11}({\bf 0},s;{\bf 0},s;\hat q)$ and its derivative.
As it is seen, the first-order correction to the energy shift vanishes, while the spin-averaged second-order term reproduces Eq.~(\ref{eq:finalresult-L}).

\bibliography{basename of .bib file}

\begin{thebibliography}{100}
	\bibitem{Agadjanov:2016cjc} 
	A.~Agadjanov, U.-G.~Mei{\ss}ner and A.~Rusetsky,
	Phys.\ Rev.\ D {\bf 95} (2017)  031502
	  [arXiv:1610.05545 [hep-lat]].

	
\bibitem{Savage:2016kon}
  M.~J.~Savage {\it et al.},
  Phys.\ Rev.\ Lett.\  {\bf 119} (2017)  062002
 [arXiv:1610.04545 [hep-lat]].

\bibitem{Chang:2015qxa}
  E.~Chang {\it et al.} [NPLQCD Collaboration],
  Phys.\ Rev.\ D {\bf 92} (2015)  114502
  [arXiv:1506.05518 [hep-lat]].


\bibitem{Beane:2014ora}
  S.~R.~Beane {\it et al.},
  Phys.\ Rev.\ Lett.\  {\bf 113} (2014)  252001
  [arXiv:1409.3556 [hep-lat]].

\bibitem{Beane:2015yha}
  S.~R.~Beane {\it et al.} [NPLQCD Collaboration],
  Phys.\ Rev.\ Lett.\  {\bf 115} (2015)  132001
  [arXiv:1505.02422 [hep-lat]].

\bibitem{Chambers:2015bka}
  A.~J.~Chambers {\it et al.},
  Phys.\ Rev.\ D {\bf 92} (2015) 114517
  [arXiv:1508.06856 [hep-lat]].

\bibitem{Chambers:2014qaa}
  A.~J.~Chambers {\it et al.} [CSSM and QCDSF/UKQCD Collaborations],
  Phys.\ Rev.\ D {\bf 90} (2014)  014510
  [arXiv:1405.3019 [hep-lat]].



	
	\bibitem{Bali:2015msa}
	G.~Bali and G.~Endr\"odi,
	Phys.\ Rev.\ D {\bf 92} (2015)  054506
  [arXiv:1506.08638 [hep-lat]].



\bibitem{Chambers:2017tuf}
  A.~J.~Chambers {\it et al.} [QCDSF and UKQCD and CSSM Collaborations],
  Phys.\ Rev.\ D {\bf 96} (2017) 114509
  [arXiv:1702.01513 [hep-lat]].


\bibitem{Chambers:2017dov}
  A.~J.~Chambers {\it et al.},
  Phys.\ Rev.\ Lett.\  {\bf 118} (2017) 242001
  [arXiv:1703.01153 [hep-lat]].





	\bibitem{Davoudi:2015cba}
	Z.~Davoudi and W.~Detmold,
	Phys.\ Rev.\ D {\bf 92} (2015)  074506
	[arXiv:1507.01908 [hep-lat]].
 


        
	\bibitem{Hagelstein:2015egb} 
	F.~Hagelstein, R.~Miskimen and V.~Pascalutsa,
	Prog.\ Part.\ Nucl.\ Phys.\  {\bf 88} (2016) 29
  [arXiv:1512.03765 [nucl-th]].

	
	\bibitem{Cottingham:1963zz}
  W.~N.~Cottingham,
  Annals Phys.\  {\bf 25} (1963) 424.

\bibitem{Carlson:2011zd}
  C.~E.~Carlson and M.~Vanderhaeghen,
  Phys.\ Rev.\ A {\bf 84} (2011) 020102
  [arXiv:1101.5965 [hep-ph]].

	
	\bibitem{Gasser:2015dwa}
	J.~Gasser, M.~Hoferichter, H.~Leutwyler and A.~Rusetsky,
	Eur.\ Phys.\ J.\ C {\bf 75} (2015) 375
	  [arXiv:1506.06747 [hep-ph]].

	\bibitem{WalkerLoud:2012bg}
	A.~Walker-Loud, C.~E.~Carlson and G.~A.~Miller,
	Phys.\ Rev.\ Lett.\  {\bf 108} (2012) 232301
	  [arXiv:1203.0254 [nucl-th]].

	\bibitem{Erben:2014hza}
	F.~B.~Erben, P.~E.~Shanahan, A.~W.~Thomas and R.~D.~Young,
	Phys.\ Rev.\ C {\bf 90} (2014)  065205
 [arXiv:1408.6628 [nucl-th]].
	

\bibitem{Cushman:2018zza}
  K.~K.~Cushman, A.~W.~Thomas and R.~D.~Young,
  arXiv:1804.05031 [nucl-th].


	\bibitem{Bernard:2002pw} 
	V.~Bernard, T.~R.~Hemmert and U.-G.~Mei{\ss}ner,
	Phys.\ Rev.\ D {\bf 67} (2003) 076008
[hep-ph/0212033].
	
	
	\bibitem{Peset:2014jxa} 
	C.~Peset and A.~Pineda,
	Nucl.\ Phys.\ B {\bf 887} (2014) 69
	 [arXiv:1406.4524 [hep-ph]].


\bibitem{Bernard:2012hb}
  V.~Bernard, E.~Epelbaum, H.~Krebs and U.-G.~Mei{\ss}ner,
  Phys.\ Rev.\ D {\bf 87} (2013) 054032
  [arXiv:1209.2523 [hep-ph]].

         
	\bibitem{Gasser:1974wd}
	J.~Gasser and H.~Leutwyler,
	Nucl.\ Phys.\ B {\bf 94} (1975) 269.

      \bibitem{Brodsky:2008qu}
  S.~J.~Brodsky, F.~J.~Llanes-Estrada and A.~P.~Szczepaniak,
  Phys.\ Rev.\ D {\bf 79} (2009) 033012
  [arXiv:0812.0395 [hep-ph]].


\bibitem{Gorchtein:2013yga}
  M.~Gorchtein, F.~J.~Llanes-Estrada and A.~P.~Szczepaniak,
  Phys.\ Rev.\ A {\bf 87} (2013)  052501
  [arXiv:1302.2807 [nucl-th]].


	\bibitem{Tarrach}
	R.~Tarrach,
	Nuovo Cim.\ A {\bf 28} (1975) 409.
	
	\bibitem{Bernabeu:1976jq}
	J.~Bernabeu and R.~Tarrach,
	Ann. Phys.\  {\bf 102} (1976) 323.
	
	\bibitem{Hill:2016bjv}
  R.~J.~Hill and G.~Paz,
  Phys.\ Rev.\ D {\bf 95} (2017) 094017
  [arXiv:1611.09917 [hep-ph]].

\bibitem{Hill:2012rh}
  R.~J.~Hill, G.~Lee, G.~Paz and M.~P.~Solon,
  Phys.\ Rev.\ D {\bf 87} (2013) 053017
  [arXiv:1212.4508 [hep-ph]].

        
	
	\bibitem{Colangelo:2006va}
	G.~Colangelo, J.~Gasser, B.~Kubis and A.~Rusetsky,
	Phys.\ Lett.\ B {\bf 638} (2006) 187  [hep-ph/0604084].
	
	
	\bibitem{Gasser:2011ju}
	J.~Gasser, B.~Kubis and A.~Rusetsky,
	Nucl.\ Phys.\ B {\bf 850} (2011) 96
  [arXiv:1103.4273 [hep-ph]].

\bibitem{Landau}
L. Landau and E. Lifshitz, \textit{Quantum Mechanics}, 3rd ed.,
Course of Theoretical Physics (Pergamon Press,
1977).	

\bibitem{IZ}
C. Itzykson and J. B. Zuber, {\it Quantum Field Theory}
(McGraw Hill, New York, 1980).

\bibitem{AS}
M. Abramowitz and I. A. Stegun, {\it Handbook of mathematical
	functions: with formulas, graphs, and mathematical
	tables} (Dover Publications, 1972).

\bibitem{Urwin}
K. Urwin,  J.
Inst. Maths. Applics. {\bf 3}, 169 (1967).

\bibitem{Meixner}
J. Meixner, F. W. Sch\"afke and G. Wolf, {\it Mathieu Functions and Spheroidal Functions and their Mathematical Foundations} (Springer-Verlag, Berlin-New York, 1980).


	\bibitem{Detmold:2006vu}
	W.~Detmold, B.~C.~Tiburzi and A.~Walker-Loud,
	Phys.\ Rev.\ D {\bf 73} (2006) 114505
          [hep-lat/0603026].


\bibitem{Hu:2007eb} 
J.~Hu, F.~J.~Jiang and B.~C.~Tiburzi,
Phys.\ Lett.\ B {\bf 653} (2007) 350
  [arXiv:0706.3408 [hep-lat]].

\bibitem{Hu:2007ts} 
J.~Hu, F.~J.~Jiang and B.~C.~Tiburzi,
Phys.\ Rev.\ D {\bf 77} (2008) 014502
  [arXiv:0709.1955 [hep-lat]].

	


      \bibitem{Tiburzi:2007ep}
  B.~C.~Tiburzi,
  Phys.\ Rev.\ D {\bf 77} (2008) 014510
  [arXiv:0710.3577 [hep-lat]].


\bibitem{Tiburzi:2008pa}
  B.~C.~Tiburzi,
  Phys.\ Lett.\ B {\bf 674} (2009) 336
  [arXiv:0809.1886 [hep-lat]].
  
\bibitem{Tiburzi:2014zva} 
B.~C.~Tiburzi,
Phys.\ Rev.\ D {\bf 89} (2014) 074019
[arXiv:1403.0878 [hep-lat]].





\bibitem{Gasser:2007zt}
  J.~Gasser, V.~E.~Lyubovitskij and A.~Rusetsky,
  Phys.\ Rept.\  {\bf 456} (2008) 167
  [arXiv:0711.3522 [hep-ph]].


\bibitem{Lozano}
  J. Lozano {\it et al.,} in progress.




      \end{thebibliography}

      \end{document}